# Reducing Multi-Dimensional Interpolation on a Grid to Quantizing the Grid Data-Base As a Recursion.

## By Roman Gitlin

### Abstract

In his article "Powerlist: A Structure for Parallel Recursion" Jayadev Misra wrote:

"Many data parallel algorithms – Fast Fourier Transform, Batcher's sorting schemes and prefix sum – exhibit recursive structure. We propose a data structure, powerlist, that permits succinct descriptions of such algorithms, highlighting the roles of both parallelism and recursion. Simple algebraic properties of this data structure can be exploited to derive properties of these algorithms and establish equivalence of different algorithms that solve the same problem."

The quote above illustrates a commonly shared assumption regarding recursion implementatations: either they are done in purely structural terms or they cannot be done at all.

Multi-dimensional interpolation on a grid is one of hosts of semi-recursive schemes that, while routinely referred to as recursive and often described in semi-recursive terms, cannot be implemented as a recursion in their structural entirety.

This article describes structural framework for and a computer implementation of a computer implemented scheme that isolates the recursive core of interpolation on a multi-grid, an arrangement that breaks down into to a number of interpolation optimization techniques which, once implemented, provide gains in multi-interpolation speed that, compared to some known benchmarks, measure in multiple orders of magnitude.

Categories and Subject Descriptors: Multi-dimensional Programming; Concurrent Programming; Recursion

# 0. Parsing path.

**Definition 0.1:** Let $Q$ be $>$-ordered hierarchy. Let $A \in .Q.$ We define node $A$ parsing closure as set $\{ S : S > A \}$.

**Definition 0.2:** Let $Q$ be a $>$-ordered hierarchy. Let $A \in Q.$ We define node $A$ parsing range as set $\{ A \} \cup \{ S \in Q : A > S \}$ that inherits hierarchy $Q$ order.

**Definition 0.3:** Let $\mathcal{A}$ be an $\succ$-ordered hierarchy. Let $X, Y \in \mathcal{A}$ be such that $X \succ Y$. We define pair $(X, Y)$ of hierarchy $\mathcal{A}$ nodes as parent / child pair if there is no $Z \in \mathcal{A}$ such that $X \succ Z \succ Y$.
∎

**Definition 0.4:** Let $\mathcal{A}$ be a $\succ$-ordered hierarchy. We define hierarchy $\mathcal{A}$ complete set of parent / child pairs as hierarchy $\mathcal{A}$ parent / child relationship.
∎

**Definition 0.5:** Let $\mathcal{A}$ be an $\succ$-ordered hierarchy. Let $^P\!\succ$ be hierarchy $\mathcal{A}$ parent / child relationship. Let relationship $^P\!\prec$ be $^P\!\succ$ reverse. We define $\succ$-ordered hierarchy $\mathcal{Q}$ as a single parent hierarchy if $^P\!\prec$ relationship is a function.
∎.

**Definition 0.6:** Let $\mathcal{A}$ be a hierarchy. We define hierarchy $\mathcal{A}$ root as hierarchy $\mathcal{A}$ maximal element.
∎.

**Definition 0.7:** Let $\mathcal{A}$ be a hierarchy. We define hierarchy $\mathcal{A}$ data-node as hierarchy $\mathcal{A}$ minimal element.
∎.

**Definition 0.8:** Let $\mathcal{A}$ be a hierarchy. We define hierarchy $\mathcal{A}$ parsing sequence as hierarchy $\mathcal{A}$ elements' strictly decreasing sequence.
∎.

**Definition 0.9:** Let $\mathcal{A}$ be a hierarchy. Let <**A**>, <**B**> be hierarchy $\mathcal{A}$ parsing sequences. We say that parsing sequence <**A**> is less than parsing sequence <**B**>, <**A**> $^{<P>}\!\prec$ <**B**>, if parsing sequence's <**A**> set of elements is a proper subset of parsing sequence's <**B**> set of elements.
∎.

**Definition 0.10:** Let $\mathcal{A}$ be a hierarchy. We define hierarchy $\mathcal{A}$ path as hierarchy $\mathcal{A}$ maximal parsing sequence.
∎.

**Theorem 0.1:** Let $\mathcal{A}$ be a hierarchy. Let <**A**> be hierarchy $\mathcal{A}$ parsing path.

Let $A_0$ be path's <**A**> first element.

Then $A_0$ is hierarchy $\mathcal{A}$ root.

**Proof:**

    Let's assume $A_0$ is not hierarchy $\mathcal{A}$ root. Then there is node $B_0 \in \mathcal{A}$ such that $B_0 \succ A_0$.

Then parsing sequence <**A**> = < $A_0$, ... > is less than parsing sequence < $B_0$, $A_0$, ... >.

**Q. E. D.**

■

**Theorem 0.2:** Let $\mathcal{A}$ be a hierarchy. Let <**A**> be hierarchy $\mathcal{A}$ path.

Let nodes **A** and **B** be path <**A**> adjacent nodes. Let **A > B**.

Then node **A** is node **B** parent.

**Proof:**

Let's assume that there is a pair **X > Y** of path < **A** > adjacent nodes such that node **X** is not node **Y** parent. Then there is node **C** ϵ <**A** > be such that **X > C > Y**.

Then node **C** can be inserted into parsing sequence <**A** > between nodes **X** and **Y** to generate a parsing sequence larger than parsing sequence <**A** >.

Then < **A** > is not hierarchy $\mathcal{A}$ path.

■

**Theorem 0.3:** Let $\mathcal{A}$ be a finite hierarchy. Let < **A** > be hierarchy $\mathcal{A}$ parsing sequence. Let nodes

**A, B** ϵ <**A**> be sequence's <**A**> first and last nodes respectively.

Then

parsing sequence  < **A** > is hierarchy $\mathcal{A}$ maximal parsing sequence joining nodes **A** and **B**

**iff**

for any pair **X, Y** of parsing sequence  < **A** > adjacent nodes such that **X > Y** it holds that node **X** is node **Y** parent.

**Proof :**

Let's assume that for any pair **X , Y** of parsing sequence < **A** > adjacent nodes such that **X > Y** it holds that node **X** is node **Y** parent.

We have to show that parsing sequence < **A** > is hierarchy $\mathcal{A}$ maximal parsing sequence joining nodes **A** and **B**.

Let's assume the opposite, namely that there is parsing sequence <  **B** > of hierarchy $\mathcal{A}$  joining nodes **A** and **B** such that  parsing sequence <**A**> is a subsequence of parsing sequence <  **B** >.

Let node **C** ϵ  < **B** > \ < **A** >. Since **C**  is  not parsing sequence's  < **A** > node, that means that

$A \neq C$ and $B \neq C$.

Since nodes **A** and **B** are sequence's $<B>$ first and last nodes respectively, that means that

$A > C > B$.

Let $A_C$ is the smallest of parsing sequence's $<A>$ nodes greater than **C**.

Then, since $A_C > C > B$, that means that node $A_C$ is not parsing sequence $<A>$ last node, and there exists node $B_C \in <A>$ such that node $B_C$ is node's $A_C$ next.

Since node $A_C$ is the smallest node greater than **C**, that means that $A_C > C > B_C$.

By the assumption, $A_C$ and $B_C$ are parsing sequence $<A>$ adjacent nodes.

Therefore node $B_C$ is node $A_C$ child.

That means that there is no node $C \in \mathcal{A}$ such that $A_C > C > B_C$.

Let's assume that parsing sequence $<A>$ is hierarchy $\mathcal{A}$ maximal parsing sequence joining nodes **A** and **B.**

Let **X, Y** be a pair of sequence's $<A>$ adjacent nodes such that $X > Y$.

We have to show that node **X** is node **Y** parent**.**

Let's assume the opposite, namely that here is node $C \in \mathcal{A}$ such that $X > C > Y$.

Then $<A>$ is not hierarchy $\mathcal{A}$ maximal parsing sequence joining nodes **A**, and **B**.

**Q.E.D**

∎

**Theorem 0.4:** Let $\mathcal{A}$ be a hierarchy. Let node $A \in <A>$. Let $\text{fl}^A$ be be node **A** parsing closure. Let $A_0$ be parsing closure's $\text{fl}^A$ maximal element. Then node $A_0$ is hierarchy's $\mathcal{A}$ root.

**Proof :**

Let's assume the opposite, namely that there is node $B \in \mathcal{A}$ such that $B > A$. Then $B \in \text{fl}^A$ and node is not parsing closure's $\text{fl}^A$ maximal element.

**Q.E.D**

# I. Meta-Parsing Hierarchies.

## 1. Meta-Parsing Hierarchy : A Generalized Parsing Scheme.

In this sub-section we define meta-parsing hierarchy – a set-theoretical template that can be instantiated as a host of quantizing-multi-array-as-a-recursion schemes in general, and quantizing-multi-array-as-a-recursion, algorithm-specific interpolation schemes in particular.

**Definition 1.1:** Let $\mathbf{Q}$ be an strictly-ordered hierarchy. We define hierarchy $\mathbf{Q}$ as a meta-parsing hierarchy as a hierarchy such that for any $\mathbf{S} \in \mathbf{Q}$ node's $\mathbf{S}$ parsing closure is a finite, linearly ordered set**.**

**Theorem 1.1 :** Let $\mathbf{Q}$ be an an $>$-ordered meta-parsing hierarchy. Hierarchy $\mathbf{Q}$ is a single parent hierarchy.

**Proof:**

> Let $\mathbf{A} \in \mathbf{Q}$. Let $\mathrm{fl}^{\mathbf{A}}$ be node $\mathbf{A}$ parsing closure. By definition, $\mathrm{fl}^{\mathbf{A}}$ is a finite linearly ordered set.
>
> Thus, there is unique node $\mathbf{S} \in \mathrm{fl}^{\mathbf{A}}$ such that node $\mathbf{S}$ is parsing closure's $\mathrm{fl}^{\mathbf{A}}$ smallest node.
>
> We have to show that node $\mathbf{S}$ is node $\mathbf{A}$ *unique* parent.
>
> Since $\mathbf{S} > \mathbf{A}$ by $\mathrm{fl}^{\mathbf{A}}$ definition, it is sufficient to prove that node $\mathbf{S}$ is hierarchy $\mathbf{Q}$ *smallest* element greater than $\mathbf{A}$**.**
>
> Let's assume that there is node $\mathbf{B} \in \mathbf{Q}$ such that $\mathbf{B}$ is not an element of $\mathrm{fl}^{\mathbf{A}}$ and $\mathbf{S} > \mathbf{B} > \mathbf{A}$.
>
> Then, by definition, $\mathbf{B} \in \mathrm{fl}^{\mathbf{A}}$.
>
> Then, node S is not parsing closure's $\mathrm{fl}^{\mathbf{A}}$ smallest element.
>
> **Q. E. D**.

**Lemma 1.1 :** Let $Q$ be an an $>$-ordered meta-parsing hierarchy. Let $A \in Q$. Let $ fl^A$ be node **a** parsing closure. Let node $A_0$ be set $fl^A$ largest node. Then node $A_0$ is node's $A$ unique ancestral root.

**Proof:**

By **theorem 0.4,** node $A_0$ is hierarchy's $Q$ root.

Let's show that node $A_0$ is node $A$ *unique* ancestral root.

Let's assume that there is hierarchy's $Q$ root $B_0$ distinct from node $A_0$ which is node $A$ ancestral root. That means that $B_0 \in fl^A$. Since $fl^A$ is a linearly ordered set, that implies that either

$A_0 > B_0$ or $B_0 > A_0$.

If $A_0 > B_0$ then node $B_0$ is not hierarchy $Q$ root.

If $B_0 > A_0$ then $A_0$ is not $fl^A$ largest element.

Q. E. D.

■

**Lemma 1.2 :** Let $Q$ be an an $>$-ordered meta-parsing hierarchy. Let $A \in Q$. Let $fl^A$ be node **a** parsing closure. Let $A_0, .., A_k$ be set's $fl^A$ totality of nodes listed in their descending order.

Then parsing sequence $< A_0, .., A_k, A >$ is hierarchy's $Q$ largest parsing sequence joining nodes $A_0$ and $A$.

**Proof:**

Let's assume that $< A_0, A_1, .., A_k, A >$ is not hierarchy $Q$ largest parsing sequence joining nodes $A_0$ and $A$. Then there is a parsing sequence $< A_0, B_1, .., B_L, A >$ such that sequence $< A_0, B_1, .., B_L, A >$ is parsing sequence's $< A_0, A_1, .., A_k, A >$ super-sequence.

That in turn means that there is node $B_j \in < A_0, B_1, .., B_L, A > \setminus < A_0, A_1, .., A_k, A >$.

Since $B_j > A$, that means that $B_j \in fl^A$.

That in turn means that $B_j \in < A_0, A_1, .., A_k, A >$.

Q. E. D.

■

**Definition 1.2 :** Let **Q** be an a **>**-ordered meta-parsing hierarchy. Let node **A** ϵ **Q**. We define node **A** ancestral path as hierarchy **Q** largest parsing sequence joining node **A** and node **A** ancestral root.
▬

**Theorem 1.2 :** Let **Q** be an an **>**-ordered meta-parsing hierarchy. Each of hierarchy's **Q** nodes uniquely defines its ancestral path.

**Proof:**

 Follows directly from **lemma 1,2**.
■

**Notation 1.1:** Let **Q** be an **>**-ordered hierarchy.

 ▼ We will be referring to parsing sequence's number of links as parsing sequence's length.

▼

**Definition 1.3:** Let **Q** be a a meta-parsing hierarchy. We define meta-parsing hierarchy **Q** level **i** node as a node whose ancestral path is of length **i**.
▬

**Lemma 1.3 :** Let **Q** be a meta-parsing hierarchy. Let **<A>** be hierarchy **Q** parsing path.

Let A ϵ **<A>**. Let **<P_A>** = < $A_0$, .., A > be path's **<A>** sub-sequence consisting of all path's **<A>** nodes preceding node **A.** Then **<P_A>** is node **A** ancestral path.

**Proof :**

 By **theorem 0.2**, since **<A>** is hierarchy **Q** parsing path, each pair of path's **<A>** adjacent nodes is in a parent /child relationship.

 By **theorem 0.3**, that means that parsing sequence < **P_A** > is hierarchy **Q** *maximal* parsing sequence joining nodes $A_0$ and **A.**

 Since, by **theorem 1.2**, there exists hierarchy **Q** largest parsing sequence joining nodes $A_0$ and **A**, hierarchy's **Q** *maximal* parsing sequence < **P_A** > joining nodes $A_0$ and **A** must be hierarchy's **Q** *largest* parsing sequence joining nodes $A_0$ and **A.**

Q. E. D.

■

**Lemma 1.3A :** Let $Q$ be a meta-parsing hierarchy. Let $A \in Q$. Let $<P_A> = <A_0, .., A>$ be node's $<A>$ ancestral path. Let node $B \in <P_A>$ such that $B > A$. Then

Let parsing sequence $<P_B> = <A_0, .., B>$ be path's $<P_A>$ sub-sequence consisting of all path's $<P_A>$ nodes preceding node $B$. Then $<P_B>$ is node's $B$ ancestral path.

**Proof :**

Analogous to proof of **lemma 1.3.**

Q. E. D.

■

**Theorem 1.3 :** Let $Q$ be a meta-parsing hierarchy. Let $<A>$ be hierarchy $Q$ parsing path. Let node $A \in <A>$. Then node $A$ is hierarchy $Q$ level $i$ node **iff** node $A$ is path $<A>$ $i^{th}$ node.

**Proof :**

Let parsing sequence $<P_A> = <A_0, .., A>$ be path's $<A>$ parsing sub-sequence consisting of all of path's $<A>$ nodes preceding node $A$.

We first observe that node $A_0$ is node's $A$ unique ancestral root ( **theorems 0.1, 0.4** and **1.2** )**.**

We next observe that, since all of parsing path's $<A>$ adjacent nodes are in a parent / child relationship, all of parsing path's $<P_A>$ adjacent nodes are in a parent / child relationship as well.

Thus, by **lemma 1.3**, parsing sequence $<P_A>$ is node's $A$ ancestral path.

Let node $A$ be hierarchy $Q$ level $i$ node.

Then, parsing sequence $<P_A>$, by virtue of being node's $A$ ancestral path, is parsing sequence of length $i$.

Therefore node $A$ is path's $<A>$ $i^{th}$ node.

Let node $A$ be path's $<A>$ $i^{th}$ node.

Since parsing sequence $<P_A>$ is node $A$ ancestral path**,** and since parsing sequence $<P_A>$ is of length $i$, node $A$ is hierarchy $Q$ level $i$ node.

Q. E. D.

∎

**Theorem 1.3A** : Let $Q$ be a meta-parsing hierarchy. Let $A \in Q$. Let $<P_A> = <A_0, .., A>$ be node's $<A>$ ancestral path. Let node $B \in <P_A>$. Then node $A$ is hierarchy $Q$ level $i$ node **iff** node $A$ is parsing sequence's $<P_A>$ $i^{th}$ node.

**Proof** :

Analogous to proof of theorem **1.3.**

Q. E. D.

∎

**Theorem 1.4** : Let $Q$ be an an $>$-ordered meta-parsing hierarchy. Let node $A \in Q$ be hierarchy terminal node. Then node $A$ uniquely defines its hierarchy's $Q$ encompassing parsing path.

**Proof:**

Let $<A>$ be node's $A$ ancestral path.

Since node $A$ uniquely defines its ancestral pat**,** it will suffice for us to show that parsing sequence $<A>$ is hierarchy $Q$ parsing path.

Let's assume that parsing sequence $<A>$ is not hierarchy $Q$ parsing path.

Then, there must be hierarchy's $Q$ parsing sequence, $<B>$**,** that is a proper super-set of parsing sequence $<A>$.

Since node $A$ is hierarchy's $Q$ minimal node, node $A$ is sequence's $<B>$ terminal node as well.

Let node $C \in <B> \setminus <A>$. That means that $C > A$.

That, in turn, means that node $C$ is an element of node's $A$ parsing closure and, **by lemma 1.2**, is an element of node's $A$ ancestral path.

Q. E. D.

∎

**Lemma 1.4** : Let $Q$ be an an $>$-ordered meta-parsing hierarchy. Hierarchy's $Q$ root is a non-empty set.

**Proof:**

Follows directly from **lemma 1.1**.

**Q. E. D**.
∎

**Theorem 1. 5 :** A meta-parsing hierarchy is a sum of its roots' parsing ranges.

**Proof:**

Let $Q$ be an an $>$-ordered meta-parsing hierarchy. Let node $A \in Q$.

By l**emma 1 / 3**, node **A** *uniquely* defines its ancestral root. Thus, hierarchy $Q$ is a *disjoint* union of its roots' parsing ranges.

**Q. E. D**.
∎

# 2. Meta-Parsing Hierarchies : Indexing Hierarchies.

In this sub-section we define indexing hierarchy – a hard wired instantiation of more general meta-parsing hierarchy.

**Definition 1.4:** Let $I^{[1/N]}_{[S_1, ..., S_N]/[s_1, ..., s_N]}$ be an indexing set. We define $[S_1, ..., S_N] / [s_1, ..., s_N]$ indexing hierarchy, $\mathcal{A}^{[1/N]}_{[S_1, ..., S_N]/[s_1, ..., s_N]}$, as a set

$$\{\ (\ )\ \} \cup \{\ I^{[1/1]}_{[S_1]/[s_1]}\} \cup \{\ I^{[1/2]}_{[S_1, S_2]/[s_1, s_2]}\} \cup ... \cup I^{[1/N]}_{[S_1, ..., S_N]/[s_1, ..., s_N]}\ \}$$

ordered as follows:

Empty string () is hierarchy's $\mathcal{A}^{[1/N]}_{[S_1, ..., S_N]/[s_1, ..., s_N]}$ root.

For $(a_1, ..., a_L)$, $(b_1, ..., b_M) \in \mathcal{A}^{[1/N]}_{[S_1, ..., S_N]/[s_1, ..., s_N]}$

$(a_1, ..., a_L) < (b_1, ..., b_M)$

    **iff**

(a) $L > M$, and

(b) $(a_1, ..., a_M) = (b_1, ..., b_M)$

▬

**Lemma 1.5:** Let $\mathcal{A}^{[1/N]}_{[S_1, ..., S_N]/[s_1, ..., s_N]}$ be an $[S_1, ..., S_N] / [s_1, ..., s_N]$ indexing hierarchy. Let

{ $A_0, ..., A_K$ } be hierarchy's $\mathcal{A}^{[1/N]}_{[S_1, ..., S_N]/[s_1, ..., s_N]}$ parsing sequence.

Then **K ≤ N.**

**Proof:**

By **definition 1.4,** for each node $A_i$, **i** = 1, ..., **K**, there is a unique natural number $L_i$ ( $L_i \leq N$ )

such that $A_i \in I^{[1/L_i]}_{[S_1, ..., S_{L_i}]/[s_1, ..., s_{L_i}]}$.

Thus, by **definition 1.4**, $L_1 < L_2 < ... < L_K$

Thus, if **K > N** then $L_K > N.$

**Q.E.D.**

∎

**Lemma 1.6:** Let $\mathcal{A}^{[1/N]}_{[S_1, ..., S_N]/[s_1, ..., s_N]}$ be an indexing hierarchy.
Let $A = (a_1, ..., a_M) \in \mathcal{A}^{[1/N]}_{[S_1, ..., S_N]/[s_1, ..., s_N]}$ and $B = (b_1, ..., b_K) \in \mathcal{A}^{[1/N]}_{[S_1, ..., S_N]/[s_1, ..., s_N]}$.
Let **A > B.** Then node **A** is node's **B** parent **iff K = M + 1**

**Proof:**

Let node $A = (a_1, ..., a_M)$ be node's $B = (b_1, ..., b_K)$ parent.

That means that **A > B.** By definition **1.4**, if node **A** is node's **B** parent, it is sufficient for us to show

that **K = M + 1.**

Let's assume that **K > M + 1.**

Since $(a_1, ..., a_M) = (b_1, ..., b_M)$, then, by **definition 1.4**,

$(a_1, ..., a_M) > (a_1, ..., a_M, b_{M+1}) > (a_1, ..., a_M, b_{M+1}, ..., b_K)$

Thus, if **K > M + 1** then, contrary to the assumption, node **A** is not node **B** parent.

Let **K = M + 1.** We have to show that node **A** is node's **B** parent.

Let's assume the opposite, namely that there is node $C = (c_1, ..., c_L) \in \mathcal{A}^{[1/N]}_{[S_1, ..., S_N]/[s_1, ..., s_N]}$ such

that $(a_1, ..., a_M) > (c_1, ..., c_L) > (a_1, ..., a_{M+1}).$

Then, by **definition 1.4**, **M < L < M + 1.**

**Q.E.D.**

**Lemma 1.7:** Let $\mathcal{A}^{[1/N]}_{[S_1, ..., S_N]/[s_1, ..., s_N]}$ be an an $[S_1, ..., S_N] / [s_1, ..., s_N]$ indexing hierarchy.

Let $A = (a_1, ..., a_L) \in \mathcal{A}^{[1/N]}_{[S_1, ..., S_N]/[s_1, ..., s_N]}$. Then

    (a) set $\{ (), (a_1), ..., (a_1, ..., a_{L-1}) \}$ is node's **A** linearly ordered parsing closure, and

    (b) parsing sequence $< (), (a_1), ..., (a_1, ..., a_{L-1}), (a_1, ..., a_L) >$ is node's **A** ancestral path.

**Proof :**

    By **definition 1.4**, parsing sequence $\{ (), (a_1), ..., (a_1, ..., a_{L-1}) \}$ is a linearly ordered set.

    By **lemma 1.6**, set $\{ (), (a_1), ..., (a_1, ..., a_{L-1}) \}$ is node's **A** parsing closure.

    By **lemma 1.2**, parsing sequence $< (), (a_1), ..., (a_1, ..., a_{L-1}), (a_1, ..., a_{L-1}) >$ is node **A** ancestral path.

**Q.E.D.**

**Theorem 1.6 :** Let $\mathcal{A}^{[1/N]}_{[S_1, ..., S_N]/[s_1, ..., s_N]}$ be an an $[S_1, ..., S_N] / [s_1, ..., s_N]$ indexing hierarchy. Hierarchy $\mathcal{A}^{[1/N]}_{[S_1, ..., S_N]/[s_1, ..., s_N]}$ is a meta-parsing hierarchy.

**Proof :**

    Follows directly from **Lemma 1.7.**

**Q.E.D.**

**Theorem 1.7 :** Let $\mathcal{A}^{[1/N]}_{[S_1, ..., S_N]/[s_1, ..., s_N]}$ be an an $[S_1, ..., S_N] / [s_1, ..., s_N]$ indexing hierarchy. Let $C^N$ be an $[S_1, ..., S_N] / [s_1, ..., s_N]$ indexing hierarchy. Let $A \in \mathcal{A}^{[1/N]}_{[S_1, ..., S_N]/[s_1, ..., s_N]}$. Then node **A** is hierarchy $\mathcal{A}^{[1/N]}_{[S_1, ..., S_N]/[s_1, ..., s_N]}$ level **L** node

        **iff**

$A \in I^{[1/L]}_{[S_1, ..., S_L]/[s_1, ..., s_L]}$.

**Proof :**

Let $A = (a_1, \ldots, a_L) \in \mathcal{A}^{[1/N]}_{[S_1, \ldots, S_N]/[s_1, \ldots, s_N]}$. Then, by **lemma 1.7,** node **A** is hierarchy's $\mathcal{A}^{[1/N]}_{[S_1, \ldots, S_N]/[s_1, \ldots, s_N]}$ level **L** node..

Let $A \in \mathcal{A}^{[1/N]}_{[S_1, \ldots, S_N]/[s_1, \ldots, s_N]}$ be hierarchy's $\mathcal{A}^{[1/N]}_{[S_1, \ldots, S_N]/[s_1, \ldots, s_N]}$ level **L** node. Let $A = (a_1, \ldots, a_M) \in \mathbf{I}^{[1/M]}_{[S_1, \ldots, S_M]/[s_1, \ldots, s_M]}$ for some $0 < M \leq N$.

We next show that **L = M.** Let's assume that $L \neq M.$

By **lemma 1.7,** $< (), (a_1), \ldots, (a_1, \ldots, a_{M-1}), (a_1, \ldots, a_M) >$ is node **A** ancestral path.

Since we assume that **A** is hierarchy $\mathcal{A}^{[1/N]}_{[S_1, \ldots, S_N]/[s_1, \ldots, s_N]}$ level **L** node, then, by definition, node **A** ancestral path $< (), (a_1), \ldots, (a_1, \ldots, a_{M-1}), (a_1, \ldots, a_M) >$ must be of length **L**.

Thus **L = M**.

**Q.E.D.**

∎

**Theorem 1.8:** Let $\mathcal{A}^{[1/N]}_{[S_1, \ldots, S_N]/[s_1, \ldots, s_N]}$ be an an $[S_1, \ldots, S_N]/[s_1, \ldots, s_N]$ indexing hierarchy. Let $A \in \mathcal{A}^{[1/N]}_{[S_1, \ldots, S_N]/[s_1, \ldots, s_N]}$ be level **i** node.

Then $i \leq N$.

**Proof:**

By definition **1. 4**, $A \in \mathbf{I}^{1/L}_{[S_1, \ldots, S]/[s_1, \ldots, s_L]}$ for some $L \leq N.$

By **theorem 1. 7**, node **A** is level **L** node.

Thus **L = i**.

**Q.E.D.**

∎

**Theorem 1.9 :** Let $\mathcal{A}^{[1/N]}_{[S_1, \ldots, S_N]/[s_1, \ldots, s_N]}$ be an an $[S_1, \ldots, S_N]/[s_1, \ldots, s_N]$ indexing hierarchy. Let $A = (a_1, \ldots, a_L) \in \mathcal{A}^{[1/N]}_{[S_1, \ldots, S_N]/[s_1, \ldots, s_N]}$ be hierarchy $\mathcal{A}^{[1/N]}_{[S_1, \ldots, S_N]/[s_1, \ldots, s_N]}$ level **L** node ( $0 \leq L < N$ ).

Then set $\{ (a_1, \ldots, a_L, a_{L+1}) : a_{L+1} \in \mathbf{I}^{L+1}_{S_{L+1}/s_{L+1}} \}$ is hierarchy's $\mathcal{A}^{[1/N]}_{[S_1, \ldots, S_N]/[s_1, \ldots, s_N]}$ $[S_{L+1}]/[s_{L+1}]$-indexed set of all of node's $(a_1, \ldots, a_L)$ children.

**Proof :**

By **lemma 1.6**, $[S_{L+1}] / [s_{L+1}]$-indexed set $\{ (a_1, \ldots, a_L, a_{L+1}) : a_{L+1} \in I^{L+1}{}_{S_{L+1}/s_{L+1}} \}$ contains all of node's **A** children.

**Q.E.D.**

∎

**Theorem 1.10 :** : Let $\mathcal{A}^{[1/N]}{}_{[S_1, \ldots, S_N]/[s_1, \ldots, s_N]}$ be an $[S_1, \ldots, S_N] / [s_1, \ldots, s_N]$ indexing hierarchy. Let $A \in \mathcal{A}^{[1/N]}{}_{[S_1, \ldots, S_N]/[s_1, \ldots, s_N]}$.
Then
  node **A** is hierarchy's $\mathcal{A}^{[1/N]}{}_{[S_1, \ldots, S_N]/[s_1, \ldots, s_N]}$ terminal node
   **iff**
  node **A** is hierarchy's $\mathcal{A}^{[1/N]}{}_{[S_1, \ldots, S_N]/[s_1, \ldots, s_N]}$ level **N** node.

**Proof :**

Let node **A** be hierarchy's $\mathcal{A}^{[1/N]}{}_{[S_1, \ldots, S_N]/[s_1, \ldots, s_N]}$ level **N** node.

By **theorem 1.7**, $A = (a_1, \ldots, a_N) \in I^{[1/N]}{}_{[S_1, \ldots, S_N]/[s_1, \ldots, s_N]}$.

Thus, **by definition 1.4,** node **A** is hierarchy's $\mathcal{A}^{[1/N]}{}_{[S_1, \ldots, S_N]/[s_1, \ldots, s_N]}$ minimal node.

Let node **A** be hierarchy's $\mathcal{A}^{[1/N]}{}_{[S_1, \ldots, S_N]/[s_1, \ldots, s_N]}$ terminal node.

Let's assume that node **A** is hierarchy's $\mathcal{A}^{[1/N]}{}_{[S_1, \ldots, S_N]/[s_1, \ldots, s_N]}$ node of order **M**.

Let's assume that $0 < M < N$.

By **theorem 1.7**, $A = (a_1, \ldots, a_M) \in I^{[1/M]}{}_{[S_1, \ldots, S_M]/[s_1, \ldots, s_M]}$.

By **theorem 1.9,** for any $a_{M+1} \in I^{M+1}{}_{S_{M+1}/s_{M+1}}$ node $(a_1, \ldots, a_M, a_{M+1})$ is node's **A** child.

**Q.E.D.**

∎

# Meta-Parsing Hierarchies : Indexing Order Hierarchies.

**Definition 1.5:** Let $\mathcal{A}, \mathcal{B}$ be hierarchies. We define hierarchies $\mathcal{A}$ and $\mathcal{B}$ as isomorphic if there is a map $\mathbf{T} : \mathcal{A} \to \mathcal{B}$ such that $\mathbf{T}$ is an onto, one-to-one, order-preserving map.

∎

**Definition 1.6 :** Let $\mathcal{B}$ be a hierarchy. We define hierarchy $\mathcal{B}$ as an $[S_1, ..., S_N] / [s_1, ..., s_N]$ *indexing order hierarchy* if hierarchy $\mathcal{B}$ is isomorphic to an $[S_1, ..., S_N] / [s_1, ..., s_N]$ *indexing hierarchy*.

∎

**Lemma 1.8:** Let $\mathcal{A}^{[1/N]}_{[S_1, ..., S_N] / [s_1, ..., s_N]}$ be indexing hierarchy. Let $\mathbf{C}^N$ be an $[S_1, ..., S_N] / [s_1, ..., s_N]$ indexing order hierarchy. Let $\mathbf{T}_{\mathbf{C}^N} : \mathcal{A}^{[1/N]}_{[S_1, ..., S_N] / [s_1, ..., s_N]} \to \mathbf{C}^N$ be an onto, one-to-one, order-preserving map. Let nodes $\mathbf{A}, \mathbf{B} \in \mathcal{A}^{[1/N]}_{[S_1, ..., S_N] / [s_1, ..., s_N]}$.

Then
    **(a)** if node **A** is node's **B** parent

        **iff**

    **(b)** node $\mathbf{T}_{\mathbf{C}^N}(\mathbf{A})$ is node's $\mathbf{T}_{\mathbf{C}^N}(\mathbf{B})$ parent

**Proof:**

    Obvious.

**Q.E.D.**
∎

**Theorem 1.11:** Indexing order hierarchy is a meta-parsing hierarchy.

**Proof:**

    Let $\mathcal{A}^{[1/N]}_{[S_1, ..., S_N] / [s_1, ..., s_N]}$ be an $[S_1, ..., S_N] / [s_1, ..., s_N]$ indexing hierarchy.

    Let $\mathbf{C}^N$ be an $[S_1, ..., S_N] / [s_1, ..., s_N]$ indexing order hierarchy.

    Let $\mathbf{T}_{\mathbf{C}^N} : \mathcal{A}^{[1/N]}_{[S_1, ..., S_N] / [s_1, ..., s_N]} \to \mathbf{C}^N$ be an onto, one-to-one, order-preserving map.

    Let a non-root node $\mathbf{A} \in \mathbf{C}^N$. Let node $\mathbf{X} = (\mathbf{T}_{\mathbf{C}^N})^{-1}(\mathbf{A})$.

Let, for some $0 < L \leq N$, $X = (c_1, ..., c_L) \in I^{[1/L]}_{[S_1, ..., S_L]/[s_1, ..., s_L]}$.

Let $< (), (c_1), ..., (c_1, ..., c_{L-1}), X >$ be node $X$ ancestral path.

By **definition 1.4**, set $\{ (), (c_1), ..., (c_1, ..., c_{L-1}) \}$ is a linearly ordered set.

By map $T_{C^N}$ definition, set $\{ T_{C^N}(()), T_{C^N}((c_1)), ..., T_{C^N}((c_1, ..., c_{L-1})) \}$ is linearly ordered set.

By **definition 1.4**, $(c_1, ..., c_i)) > (c_1, ..., c_L))$ for $i = 1, ..., L - 1$.

By map $T_{C^N}$ monotonicity, $T_{C^N}((c_1, ..., c_i)) > T_{C^N}((c_1, ..., c_L)) =$ for $i = 1, ..., L - 1$.

Thus, set $\{ T_{C^N}(()), T_{C^N}((c_1)), ..., T_{C^N}((c_1, ..., c_{L-1})) \}$ is a subset of node $T_{C^N}(X)$ parsing closure.

By **lemma 1.6**, node $(c_1, ..., c_i)$ is node's $(c_1, ..., c_{i-1})$ child for $i = 1, ..., L - 1.$.

By **lemma 1. 8**, node $T_{C^N}((c_1, ..., c_i))$ is node's $T_{C^N}((c_1, ..., c_{i-1}))$ child for $i = 1, ..., L - 1.$

Thus, all of parsing sequence

$\{ T_{C^N}(()), T_{C^N}((c_1)), ..., T_{C^N}((c_1, ..., c_{L-1}))) \}$

adjacent nodes are in a parent / child relationship.

By **theorem 0.3**, parsing sequence

$<T_{C^N}(()), T_{C^N}((c_1)), ..., T_{C^N}((c_1, ..., c_{L-1})), T_{C^N}((c_1, ..., c_L)) >$

is hierarchy's $C^N$ maximal paesing sequence joining nodes $T_{C^N}(())$ and $T_{C^N}((c_1, ..., c_L))$..

Thus set $\{ T_{C^N}(()), T_{C^N}((c_1)), ..., T_{C^N}((c_1, ..., c_{L-1})) \}$ is node's $T_{C^N}((c_1, ..., c_L))$ linearly ordered, finite parsing closure.

Since, by definition, $T_{C^N}$ is an onto map, hierarchy is a meta-parsing hierararchy.

**Q.E.D.**

∎

**Lemma 1.9:** Let $\mathcal{A}^{[1/N]}_{[S_1, ..., S_N]/[s_1, ..., s_N]}$ be an $[S_1, ..., S_N] / [s_1, ..., s_N]$ indexing hierarchy. Let $C^N$ be an $[S_1, ..., S_N] / [s_1, ..., s_N]$ indexing order hierarchy. Let $T_{C^N} : \mathcal{A}^{[1/N]}_{[S_1, ..., S_N]/[s_1, ..., s_N]} \to C^N$ be an onto, one-to-one, order-preserving map.

Let $A \in \mathcal{A}^{[1/N]}_{[S_1, ..., S_N]/[s_1, ..., s_N]}$.

Then node $T_{C^N}(A) \in C^N$ is hierarchy's $C^N$ level $L$ node

                **iff**

node $A$ is hierarchy $\mathcal{A}^{[1/N]}_{[S_1, ..., S_N]/[s_1, ..., s_N]}$ level $L$ node

**Proof:**

    By **theorem 1.7**, node $A$ is hierarchy $\mathcal{A}^{[1/N]}_{[S_1, ..., S_N]/[s_1, ..., s_N]}$ level $L$ node

    **iff** $A = (a_1, ..., a_L) \in I^{[1/L]}_{[S_1, ..., S_L]/[s_1, ..., s_L]}$.

    Thus all we have to prove is is that

    node $T_{C^N}(A) \in C^N$ is hierarchy's $C^N$ level $L$ node

                **iff**

    node $A = (a_1, ..., a_L) \in I^{[1/L]}_{[S_1, ..., S_L]/[s_1, ..., s_L]}$.

    Let node $A = (a_1, ..., a_L) \in I^{[1/L]}_{[S_1, ..., S_L]/[s_1, ..., s_L]}$.

    Then, set $\{ \,(), (a_1), ..., (a_1, ..., a_{L-1}) \,\}$ is node $A$ parsing closure.

    Then $\{ T_{C^N}(()), T_{C^N}((a_1)), ..., T_{C^N}((a_1, ..., a_{L-1})) \}$ is node's $T_{C^N}(A)$ parsing closure (see **theorem 1.11** proof ).

    The, by **lemma 1.2**,

    parsing sequence $< T_{C^N}(()), T_{C^N}((a_1)), ..., T_{C^N}((a_1, ..., a_{L-1})), T_{C^N}((a_1, ..., a_L)) >$ is

    node's $T_{C^N}(A)$ ancestral path of length N.

    Thus, node $T_{C^N}(A)$ is hierarchy's $C^N$ level $L$ node.

Let node $B \in C^N$ be hierarchy's $C^N$ level $L$ node.

Let parsing sequence $\{B_0, B_1, ..., B_{L-1}\}$ be node $B$ parsing closure.

By map $T_{C^N}$ monotonicity, set $\{(T_{C^N})^{-1}(B_0), (T_{C^N})^{-1}(B_1), ..., (T_{C^N})^{-1}(B_{L-1})\}$ is a subset of node $(T_{C^N})^{-1}(B)$ parsing closure.

By **lemma 1.8**, set $\{(T_{C^N})^{-1}(B_0), (T_{C^N})^{-1}(B_1), ..., (T_{C^N})^{-1}(B_{L-1})\}$ is node $(T_{C^N})^{-1}(B)$ parsing closure.

By **lemma 1/2**, $<(T_{C^N})^{-1}(B_0), (T_{C^N})^{-1}(B_1), ..., (T_{C^N})^{-1}(B_{L-1}), (T_{C^N})^{-1}(B)>$ is node's $(T_{C^N})^{-1}(B)$ ancestral path.

Thus node $(T_{C^N})^{-1}(B)$ is hierarchy's $\mathcal{A}^{[1/N]}_{[S_1, ..., S_N]/[s_1, ..., s_N]}$ level $L$ node.

**Q.E.D.**

∎

**Theorem 1.12:** Let $C^N$ be an $[S_1, ..., S_N] / [s_1, ..., s_N]$ indexing order hierarchy.
Let $L < N$. Let node $X \in C^N$ be hierarchy's $C^N$ level $L$ node..
Then node $X$ has $S_{L+1} [S_{L+1}] / [s_{L+1}]$-indexed children.

**Proof :**

Let $\mathcal{A}^{[1/N]}_{[S_1, ..., S_N]/[s_1, ..., s_N]}$ be an $[S_1, ..., S_N] / [s_1, ..., s_N]$ indexing hierarchy.

Let $T_{C^N} : \mathcal{A}^{[1/N]}_{[S_1, ..., S_N]/[s_1, ..., s_N]} \to C^N$ be an onto, one-to-one, order-preserving map.

Let $L < N$. Let node $X \in C^N$ be hierarchy's $C^N$ level $L$ node..

By definizion, $T_{C^N}$ is an onto map.

Terefore there is node $A \in \mathcal{A}^{[1/N]}_{[S_1, ..., S_N]/[s_1, ..., s_N]}$ such that $X = T_{C^N}(A)$.

By **lemma 1.9**, since node **X** is hierarchy's $C^N$ level **L** node, node **A** is hierarchy's

$\mathcal{A}^{[1/N]}_{[S_1, ..., S_N]/[s_1, ..., s_N]}$ level **L** node as well.

By **theorem 1.7**, node $A = (a_1, ..., a_L) \in I^{[1/L]}_{[S_1, ..., S_L]/[s_1, ..., s_N]}$.

By **theorem 1.9**, set $\{ (a_1, ..., a_L, a_{L+1}) : a_{L+1} \in I^{L+1}_{S_{L+1}/s_{L+1}} \}$ is node's **A** total

$[S_{L+1}] / [s_{L+1}]$-indexed set of children.

Then, by **theorem 1.12** set $\{ T_{C^N} ( ( a_1, ..., a_L, a_{L+1} ) ) : a_{L+1} \in I^{L+1}_{S_{L+1}/s_{L+1}} \}$ is hierarchy's

$C^N [S_{L+1}] / [s_{L+1}]$-indexed set of all of node $T_{C^N} ( A )$ children.

**Q.E.D.**

∎

**Theorem 1.13:** Let $\mathcal{A}^{[1/N]}_{[S_1, ..., S_N]/[s_1, ..., s_N]}$ be an indexing hierarchy.

Let $C^N$ be an $[S_1, ..., S_N] / [s_1, ..., s_N]$ indexing order hierarchy.

Let $T_{C^N} : \mathcal{A}^{[1/N]}_{[S_1, ..., S_N]/[s_1, ..., s_N]} \to C^N$ be an onto, one-to-one, order-preserving map.

Then
    node $A \in \mathcal{A}^{[1/N]}_{[S_1, ..., S_N]/[s_1, ..., s_N]}$ is hierarchy $\mathcal{A}^{[1/N]}_{[S_1, ..., S_N]/[s_1, ..., s_N]}$ terminal node

              **iff**

node $T_{C^N} ( A )$ is hierarchy's $C^N$ terminal node.

**Proof :**

By **theorem 1.10**, node $A \in \mathcal{A}^{[1/N]}_{[S_1, ..., S_N]/[s_1, ..., s_N]}$ is hierarchy $\mathcal{A}^{[1/N]}_{[S_1, ..., S_N]/[s_1, ..., s_N]}$
terminal node **iff** node $A = (a_1, ..., a_N) \in I^{[1/N]}_{[S_1, ..., S_N]/[s_1, ..., s_N]}$

Thus, to prove the theorem, it is sufficient for us to show that

$T_{C^N} ( A ) \in C^N$ is hierarchy's $C^N$ terminal node

**iff**

node $A = (a_1, \ldots, a_N) \in I^{[1/N]}_{[S_1, \ldots, S_N]/[s_1, \ldots, s_N]}$.

Let node $A = (a_1, \ldots, a_N) \in I^{[1/N]}_{[S_1, \ldots, S_N]/[s_1, \ldots, s_N]}$.

Then, set $\{ \,(), (a_1), \ldots, (a_1, \ldots, a_{N-1}) \,\}$ is node **A** parsing closure.

Then, , set $\{ T_{C^N}(\,(\,)\,), T_{C^N}((a_1)), \ldots, T_{C^N}(\,(a_1, \ldots, a_{N-1})\,) \}$ is node's $T_{C^N}(A)$ parsing closure (see **theorem 1.11** proof ).

In order to show that node's $T_{C^N}(\,(a_1, \ldots, a_N)\,)$ is hierarchy's $C^N$ terminal node, we have to show that hierarchy's $C^N$ parsing sequence

$\{ T_{C^N}(\,(\,)\,), T_{C^N}((a_1)), \ldots, T_{C^N}(\,(a_1, \ldots, a_{N-1})\,), T_{C^N}(\,(a_1, \ldots, a_{N-1})\,) \}$

is hierarchy's $C^N$ maximal parsing sequence.

Let's assume that set $\{ T_{C^N}(\,(\,)\,), T_{C^N}(\,(a_1)\,), \ldots, T_{C^N}(\,(a_1, \ldots, a_{N-1})\,) \}$ is not hierarchy's $C^N$ maximal parsing sequence.

Then, there is hierarchy's $C^N$ parsing sequence $\{ X_0, \ldots, X_K \}$ – a proper superset of parsing sequence $\{ T_{C^N}(\,(\,)\,), T_{C^N}(\,(a_1)\,), \ldots, T_{C^N}(\,(a_1, \ldots, a_{N-1})\,) \}$.

As such, parsing sequence $\{ X_0, \ldots, X_K \}$ would be of length greater than **N**.

Then, hierarchy's $\mathcal{A}^{[1/N]}_{[S_1, \ldots, S_N]/[s_1, \ldots, s_N]}$ parsing sequence

$\{ (T_{C^N})^{-1}(X_0), \ldots, (T_{C^N})^{-1}(X_K) \}$ would be of length greater than **N**.

That contradicts **Lemma 1 / 5** conclusion.

Let's now show that if node $A \in C^N$ is hierarchie's $C^N$ minimal / terminal node, then node $X = (T_{C^N})^{-1}(A)$ is hierarchie's $\mathcal{A}^{[1/N]}_{[S_1, \ldots, S_N]/[s_1, \ldots, s_N]}$ minimal / terminal node.

Let's assume that node **X** is not hierarchie's $\mathcal{A}^{[1/N]}_{[S_1, \ldots, S_N]/[s_1, \ldots, s_N]}$ minimal node.

Then, by **theorem 1 /10**, $X = (a_1, \ldots, a_M) \in I^{[1/M]}_{[S_1, \ldots, S_M]/[s_1, \ldots, s_M]}$ for some $M < N$.

Then, by **lemma 1.8,** for any node $(a_1, \ldots, a_M, a_{M+1}) \in I^{[1/M+1]}_{[S_1, \ldots, S_{M+1}]/[s_1, \ldots, s_{M+1}]}$

$T_{C^N}( (a_1, …, a_M, a_{M+1}) )$ is node's $T_{C^N} ( (a_1, …, a_M) )$ child.

Thus, contrary to the assumption, node's $T_{C^N}( (a_1, …, a_M) )$ is not hierarchie's $C^N$ minimal / terminal node

**Q.E.D.**

∎

# II. Multi-Cube.

In this section we define multi-cube – the meta-parsing-hierarchy's instantiation that, as a structural template, provides a framework for recursively quantizing multi-arrays in multiple dimensions.

## 1. Multi-Cube : Definition.

**Definition 2.1:** Let be **N** a positive natural number. Let $S_1, …, S_N$ be **N** positive natural numbers and let $s_1, …, s_N$ be **N** integer numbers..

We define $[S_1, …, S_N] / [s_1, …, s_N]$ multi-cube $C^N$ as a single root meta-parsing hierarchy such that

(a) Each of the hierarchy's level **i** nodes ( **i < N** ) has $S_{i+1}$ children.

(b) Each of the hierarchy's level **i** sibling sets ( **0 < i ≤ N** ) is an $[S_i] / [s_i]$-indexed set.

(c) Each of the hierarchy's level **N** nodes is a minimal / terminal / data node.

▬

## 1. Multi-Cube : General Properties.

**Lemma 2.1:** Let $C^N$ be an $[S_1, …, S_N] / [s_1, …, s_N]$ multi-cube. Let **A** ∈ $C^N$. Let $A_0$ be multi-cube $C^N$ root. Let $<P_A> = <A_0, …, A>$ be multi-cube $C^N$ parsing sequence joining root $A_0$ and node **A**. Then $<P_A>$ is node **A** ancestral path

*iff*

each pair of parsing sequence's $<P_A>$ adjacent nodes is in multi-cube $C^N$ parent / child relationship.

**Proof:**

Let each pair of parsing sequence's <**P_A**> adjacent nodes be in multi-cube $C^N$ parent / child relationship.

Then, by **theorem 0.3**, multi-cube's $C^N$ parsing sequence <**P_A**> is multi-cube's $C^N$ maximal parsing sequence joining nodes $A_0$ and $A$.

Then, by **lemma 1.2,** if parsing sequence <**P_A**> is maximal parsing sequence joining nodes $A_0$ and $A$ then parsing sequence <**P_A**> is greatest parsing sequence joining nodes $A_0$ and $A$.

The reverse is true by default : if parsing sequence <**P_A**> is greatest parsing sequence joining nodes $A_0$ and $A$, then parsing sequence <**P_A**> is also maximal parsing sequence joining nodes $A_0$ and $A$.

**Q.E.D.**

**Theorem 2.1:** Let $C^N$ be a multi-cube. Let <**A**> be multi-cube $C^N$ parsing path. Then each pair of path's <**A**> adjacent nodes is in multi-cube $C^N$ parent / child relationship.

**Proof:**

Let $A_0$ be multi-cube $C^N$ root. Let $A$ be parsing sequence <**A**> terminal node.
Multi-cube $C^N$ parsing path <**A**> is the largest parsing sequence joining nodes $A_0$ and $A$.
Thus, each pair of path's <**A**> adjacent nodes is in multi-cube $C^N$ parent / child relationship.

**Q.E.D.**

## 3. Multi-Cube : Indexing Order Hierarchy.

**Definition 2.2 :** Let $C^N$ be $[S_1, …, S_N] / [s_1, …, s_N]$ multi-cube. Let $A_0$ be multi-cube's $C^N$ root. Let $\mathcal{A}^{[1/N]}_{[S_1, …, S_N]/[s_1, …, s_N]}$ be an indexing hierarchy. We define multi-cube's $C^N$ reverse indexing map,

$R_{C^N, } : C^N \to \mathcal{A}^{[1/N]}_{[S_1, …, S_N]/[s_1, …, s_N]}$ as follows:

For multi-cube's $C^N$ root $A_0$ we define $R_{C^N}(A_0)$ as hierarchy's $\mathcal{A}^{[1/N]}_{[S_1, …, S_N]/[s_1, …, s_N]}$

empty string.

Let node $A \in C^N$ be multi-cube's $C^N$ level $L$ node.

Node $A$ uniquely defines its ancestral path, $<P_A>$, of length $L$.

By **definition 2.1** and **theorem 1.3A,** parsing sequence $<P_A>$ is of length $\leq N$.

Let $<P_A>$ $i^{th}$ nodes be $a_i$-indexed within their encompassing sibling set, $i = 1, \ldots, L$.

We define $R_{C^N}(A)$ as node $(a_1, \ldots, a_L) \in \mathcal{A}^{[1/N]}_{[S_1, \ldots, S_N]/[s_1, \ldots, s_N]}$.

---

**Lemma 2.2:** Let $C^N$ be $[S_1, \ldots, S_N] / [s_1, \ldots, s_N]$ multi-cube. Let $\mathcal{A}^{[1/N]}_{[S_1, \ldots, S_N]/[s_1, \ldots, s_N]}$ be an $[S_1, \ldots, S_N] / [s_1, \ldots, s_N]$ indexing hierarchy. Let map $R_{C^N}: C^N \to \mathcal{A}^{[1/N]}_{[S_1, \ldots, S_N]/[s_1, \ldots, s_N]}$ be multi-cube's $C^N$ reverse indexing map. Then map $R_{C^N}$ is an onto, one-to-one map.

**Proof:**

Clearly, map $R_{C^N}$ is a one-to-one map.

Since multi-cube $C^N$ and indexing hierarchy $\mathcal{A}^{[1/N]}_{[S_1, \ldots, S_N]/[s_1, \ldots, s_N]}$ share indexing set $T^{[1/N]}_{[S_1, \ldots, S_N]/[s_1, \ldots, s_N]}$, for any node $(a_1, \ldots, a_L) \in \mathcal{A}^{[1/N]}_{[S_1, \ldots, S_N]/[s_1, \ldots, s_N]}$ we can generate a parsing sequence $<A_0, \ldots, A_i, A_{i+1}, \ldots, A_L>$ such that $A_i$ is multi-cube's $C^N$ level $i$ node, and $A_{i+1}$ is node's $A_i$ child that is $a_{i+1}$-indexed within its encompassing sibling set, $i = 0, \ldots, L-1$.

By **theorem 0.3**, parsing sequence $<A_0, \ldots, A_i, A_{i+1}, \ldots, A_L>$ is node's $A_L$ ancestral path.

By $R_{C^N}$ definition, $R_{C^N}(A_L) = (a_1, \ldots, a_L)$.

Thus, $R_{C^N}(C^N) = \mathcal{A}^{[1/N]}_{[S_1, \ldots, S_N]/[s_1, \ldots, s_N]}$.

**Q.E.D.**

∎

**Notation 2.1:** Let $C^N$ be $[S_1, \ldots, S_N] / [s_1, \ldots, s_N]$ multi-cube. Let $\mathcal{A}^{[1/N]}_{[S_1, \ldots, S_N]/[s_1, \ldots, s_N]}$ be an

$[S_1, …, S_N] / [s_1, …, s_N]$ indexing hierarchy. Let map $\mathbf{R}_{\mathcal{C}^N}: \mathcal{C}^N \to \mathcal{A}^{[1/N]}{}_{[S_1, …, S_N] / [s_1, …, s_N]}$ be multi-cube's $\mathcal{C}^N$ reverse indexing map.

Let node $\mathbf{A} \in \mathcal{C}^N$ be multi-cube $\mathcal{C}^N$ level **i** node such that $\mathbf{R}_{\mathcal{C}^N}(\mathbf{A}) = .(\mathbf{a}_1, …, \mathbf{a}_i)$.

▼ With no ambiguity arising, will be referring to multi-cube's $\mathcal{C}^N$ node **A** as multi-cube's $\mathcal{C}^N$ $[\mathbf{a}_1, …, \mathbf{a}_i]$ node.

▼

**Lemma 2. 3 :** Let $\mathcal{C}^N$ be an $[S_1, …, S_N] / [s_1, …, s_N]$ multi-cube. Let $\mathcal{A}^{[1/N]}{}_{[S_1, …, S_N] / [s_1, …, s_N]}$ be an $[S_1, …, S_N] / [s_1, …, s_N]$ indexing hierarchy. Let map $\mathbf{R}_{\mathcal{C}^N}: \mathcal{C}^N \to \mathcal{A}^{[1/N]}{}_{[S_1, …, S_N] / [s_1, …, s_N]}$ be multi-cube's $\mathcal{C}^N$ reverse indexing map.

Map $\mathbf{R}_{\mathcal{C}^N}$ is an order preserving map.

**Proof:**

Let **L, M** be natural numbers such that $0 \leq \mathbf{L}, \mathbf{M} \leq \mathbf{N}$.

Let $(\mathbf{a}_1, …, \mathbf{a}_L), (\mathbf{b}_1, …, \mathbf{b}_M) \in \mathcal{A}^{[1/N]}{}_{[S_1, …, S_N] / [s_1, …, s_N]}$.

Let nodes $[\mathbf{a}_1, …, \mathbf{a}_M], [\mathbf{b}_1, …, \mathbf{b}_L] \in \mathcal{C}^N$.

Let $[\mathbf{a}_1, …, \mathbf{a}_M] > [\mathbf{b}_1, …, \mathbf{b}_L]$.

Let nodes $(\mathbf{b}_1, …, \mathbf{b}_L), (\mathbf{b}_1, …, \mathbf{b}_L, …, \mathbf{b}_N) \in \mathcal{A}^{[1/N]}{}_{[S_1, …, S_N] / [s_1, …, s_N]}$.

Let node $[\mathbf{b}_1, …, \mathbf{b}_L] = \mathbf{R}_{\mathcal{C}^N}(\mathbf{R}_{\mathcal{C}^N})^{-1}((\mathbf{b}_1, …, \mathbf{b}_L))$

Let node $[\mathbf{b}_1, …, \mathbf{b}_L, …, \mathbf{b}_N] = \mathbf{R}_{\mathcal{C}^N}(\mathbf{R}_{\mathcal{C}^N})^{-1}((\mathbf{b}_1, …, \mathbf{b}_L, …, \mathbf{b}_N))$

Then, by $\mathbf{R}_{\mathcal{C}^N}$ definition, parsing sequence $<\mathbf{A}> = <\mathbf{b}_1, …, \mathbf{b}_L, …, \mathbf{b}_N>$ is node's $[\mathbf{b}_1, …, \mathbf{b}_L, …, \mathbf{b}_N]$ ancestral path, and parsing sequence $<\mathbf{B}> = <\mathbf{b}_1, …, \mathbf{b}_L>$ is node's $[\mathbf{b}_1, …, \mathbf{b}_L]$ ancestral path.

Then, node [ $b_1, ..., b_L$ ] ϵ < $b_1, ..., b_L, ..., b_N$ >.

Then, since [ $a_1, ..., a_M$ ] > [ $b_1, ..., b_L$ ], and since parsing sequence < $b_1, ..., b_L$ > is hierarchy's largest parsing sequence joining multi-cube's $C^N$ root and node [ $a_1, ..., a_M$ ], node [ $a_1, ..., a_M$ ] ϵ < $b_1, ..., b_L$ >.

Then, since path < $a_1, ..., a_M$ > is node's [ $a_1, ..., a_M$ ] ancestral path, and since [ $a_1, ..., a_M$ ] > [ $b_1, ..., b_L$ ], path < $a_1, ..., a_M$ > is a proper sub-sequence of path < $b_1, ..., b_L$ >.
Thus,
  (a) $M < L$, and
  (b) ( $a_1, ..., a_M$ ) = ( $b_1, ..., b_M$ ).

Thus, ( $a_1, ..., a_M$ ) > ( $b_1, ..., b_L$ )

   Q.E.D.
∎

**Definition 2.3** : Let $C^N$ be [$S_1, ..., S_N$] / [$s_1, ..., s_N$] multi-cube. Let $A_0$ be multi-cube's $C^N$ root. Let $\mathcal{A}^{[1/N]}_{[S_1, ..., S_N]/[s_1, ..., s_N]}$ be an [$S_1, ..., S_N$] / [$s_1, ..., s_N$] indexing hierarchy.

We define multi-cube's $C^N$ indexing map, $T_{C^N}$, : $\mathcal{A}^{[1/N]}_{[S_1, ..., S_N]/[s_1, ..., s_N]} \rightarrow C^N$ as $(R_{C^N})^{-1}$

—

**Lemma 2. 4 :** T is an onto, one-to-one, order-preserving map.

**Proof:**
   Obvious.

   Q.E.D.
∎

**Theorem 2.2 :** Let $C^N$ be a hierarchy.

Hierarchy $C^N$ is an [$S_1, ..., S_N$] / [$s_1, ..., s_N$] multi-cube

      iff

hierarchy $C^N$ is an [$S_1, ..., S_N$] / [$s_1, ..., s_N$] indexing order hierarchy.

**Proof :**

We have shown that if hierarchy $C^N$ is an $[S_1, ..., S_N] / [s_1, ..., s_N]$ multi-cube then hierarchy $C^N$ is an $[S_1, ..., S_N] / [s_1, ..., s_N]$ indexing order hierarchy.

Let hierarchy $C^N$ be $[S_1, ..., S_N] / [s_1, ..., s_N]$ indexing order hierarchy.

Then, by **theorems 1.6** hierarchy $C^N$ is a meta-parsing hierarchy,

by **theorems 1.9,** each of the hierarchy's level **i** ( **i** < **N** ) nodes has $S_{i+1}$

$[S_{i+1}] / [s_{i+1}]$-indexed children, and

by **theorems 1.10,** each of the hierarchy's $C^N$ level **N** nodes is a data node.

**Q.E. D.**

∎

## Multi-Cube: Parsing Ranges.

**Definition 2.3:** Let $C^N$ be $[S_1, ..., S_N] / [s_1, ..., s_N]$ multi-cube. Let $I^{[1/N]}_{[S_1, ..., S_N]/[s_1, ..., s_N]}$ be multi-cube $C^N$ indexing set. We define multi-cube $C^N$ depth **0** parsing range as multi-cube $C^N$.
▬

**Definition 2.4:** Let **M** be a natural number, $0 < M < N$. Let $C^N$ be an $[S_1, ..., S_N] / [s_1, ..., s_N]$ multi-cube. Let $I^{[1/N]}_{[S_1, ..., S_N]/[s_1, ..., s_N]}$ be multi-cube $C^N$ indexing set.
Let $(a_1, ..., a_M) \in I^{[1/M]}_{[S_1, ..., S_M]/[s_1, ..., s_M]}$. We define multi-cube's $C^N$ depth **M** $(a_1, ..., a_M)$-parsing range, $C^{N-M}_{[a_1, ..., a_M]}$, as multi-cube $C^N$ node's $[a_1, ..., a_M]$ parsing range.
▬

**Definition 2.5:** Let $C^N$ be $[S_1, ..., S_N] / [s_1, ..., s_N]$ multi-cube. Let $I^{[1/N]}_{[S_1, ..., S_N]/[s_1, ..., s_N]}$ be multi-cube $C^N$ indexing set. Let $(a_1, ..., a_N) \in I^{[1/N]}_{[S_1, ..., S_N]/[s_1, ..., s_N]}$. We define multi-cube $C^N$ depth **N** parsing range, $C^0_{[a_1, ..., a_N]}$, as multi-cube $C^N$ data-node $[a_1, ..., a_N]$.
▬

**Lemma 2.3 :** Let $C^N$ be $[S_1, ..., S_N] / [s_1, ..., s_N]$ multi-cube. Let $I^{[1/N]}_{[S_1, ..., S_N]/[s_1, ..., s_N]}$ be multi-cube $C^N$ indexing set. Let $(a_1, ..., a_M) \in I^{[1/M]}_{[S_1, ..., S_M]/[s_1, ..., s_M]}$. Let $C^{N-M}_{[a_1, ..., a_M]}$ be multi-cube $C^N$ depth **M** parsing range. Let $A \in C^{N-M}_{[a_1, ..., a_M]}$. Let.
Then node **A** is multi-cube $C^N$ level **i** node **iff** node **A** is hierarchy's $C^{N-M}_{[a_1, ..., a_M]}$ level **i** – **M** node.

**Proof :**

Let node **A** be multi-cube $C^N$ level **i** node.

Let $\text{fl}^A = \{A_0, A_1, \ldots, [a_1, \ldots, a_M], \ldots, A_i\}$ be node **A** parsing closure withing multi-cube $C^N$, $\text{fl}^A$ elements being listed in their descending order.

Then, by **theorem 2 / 4**, $\text{fl}^A = \{A_0, [a_1], \ldots, [a_1, \ldots, a_M], \ldots, [a_1, \ldots, a_M, \ldots, a_i]\}$.

By **Lemma 1.3**, within multi-cube $C^N$, parsing sequence

$\{A_0, [a_1], \ldots, [a_1, \ldots, a_M], \ldots, [a_1, \ldots, a_M, \ldots, a_{M+1}], A\}$ is largest parsing sequence joining nodes $A_0$ and **A.**

Thus parsing sequence $\{[a_1, \ldots, a_M], \ldots, [a_1, \ldots, a_M, \ldots, a_i], A\}$ is multi-cube's $C^N$ largest parsing sequence joining nodes $[a_1, \ldots, a_M]$ and **A**
.

Thus parsing sequence $\{[a_1, \ldots, a_M], \ldots, [a_1, \ldots, a_M, \ldots, a_i], A\}$ is hierarchy's $C^{N-M}{}_{[a_1, \ldots, a_M]}$ largest parsing sequence joining nodes $[a_1, \ldots, a_M]$ and **A.**

Thus node **A** is hierarchy's $C^{N-M}{}_{[a_1, \ldots, a_M]}$ level **i – M** node.

Let node **A** be hierarchy's $C^{N-M}{}_{[a_1, \ldots, a_M]}$ level **i – M** node.

Then parsing sequence $\{[a_1, \ldots, a_M], \ldots, [a_1, \ldots, a_M, \ldots, a_{i-1}], A\}$ is hierarchy's $C^{N-M}{}_{[a_1, \ldots, a_M]}$ largest parsing sequence joining nodes $[a_1, \ldots, a_M]$ and **A.**

Then parsing sequence $\{[a_1, \ldots, a_M], \ldots, [a_1, \ldots, a_M, \ldots, a_{i-1}], A\}$ is multi-cube's $C^N$ largest parsing sequence joining nodes $[a_1, \ldots, a_M]$ and **A.**

By **Lemma 1.3**, $\{A_0, [a_1], \ldots, [a_1, \ldots, a_M]\}$ is multi-cube's $C^N$ largest parsing sequence joining nodes $A_0$ and $[a_1, \ldots, a_M]$.

Thus, $\{A_0, [a_1], \ldots, [a_1, \ldots, a_M], \ldots, [a_1, \ldots, a_M, \ldots, a_{i-1}], A\}$ is multi-cube's $C^N$ largest parsing sequence joining nodes $A_0$, and **A.**

Thus, node **A** is multi-cube's $C^N$ level **i** node.

**Q.E.D.**

∎

**Theorem 2.7 :** Let $C^N$ be $[S_1, ..., S_N] / [s_1, ..., s_N]$ multi-cube. Let $I^{[1/N]}_{[S_1, ..., S_N]/[s_1, ..., s_N]}$ be multi-cube $C^N$ indexing set. Let $(a_1, ..., a_M) \in I^{[1/M]}_{[S_1, ..., S_M]/[s_1, ..., s_M]}$. Let $C^{N-M}_{[a_1, ..., a_M]}$ be multi-cube $C^N$ depth $M$ parsing range.

Then $C^{N-M}_{[a_1, ..., a_M]}$ is multi-cube's $C^N$ $[S_{M+1}, ..., S_N] / [s_{M+1}, ..., s_N]$ sub-cube.

**Proof :**

> By definition, hierrarchy $C^{N-M}_{[a_1, ..., a_M]}$ is a subhierarchy of hierarchy $C^N$. Therefore $C^{N-M}_{[a_1, ..., a_M]}$ is a meta-parsing hierarchy.
> By definition, hierrarchy $C^{N-M}_{[a_1, ..., a_M]}$ level $N - M$ node is multi-cube $C^N$ level $N$ node. Thus each of hierarchy $C^{N-M}_{[a_1, ..., a_M]}$ level $N - M$ node is hierarchy $C^{N-M}_{[a_1, ..., a_M]}$ terminal node.
>
> By **Lemma 2 / 2,** each sibling set of hierarchy $C^{N-M}_{[a_1, ..., a_M]}$ level $i$ nodes is a sibling set of multi-cube $C^N$ level $i + M$ nodes ( $0 < i \leq N - M$ ).
>
> **Q.E.D.**

∎

**Notation 2.3:** Let $C^N$ be a multi-cube.

> ▼ We will be referring to multi-cube's $C^N$ as a multi-cube of type $P$ if multi-cube $C^N$ data-set elements are of type $P$..

▼

# III. Multi-Array.

All of subscript operators share a taken for granted flaw: *in order to be used, a subscript operator must be mplemented first.*

One of the structural statements this article makes – and its accompanying code *implements* – is a rejection of reliance on subscript operator either as heuristic means of multi-array description or as multi-array parsing means.

In this section we derive multi-array basic properties in set-theoretical, non-subscript, *non-heuristic* terms.

**Definition 3.1 :** Let $Q^N = TI^{[1/N]}_{[S_1, ..., S_N]/[s_1, ..., s_N]}$ be an array.

We define multi-array $Q^N$ Cartesian Indexing order, $\prec^{Q^N}_{[S_1, ..., S_N]/[s_1, ..., s_N]}$, as follows :

For $((b_1, ..., b_N), Q^N((b_1, ..., b_N)))$ and $((c_1, ..., c_N), Q^N((c_1, ..., c_N))) \in Q^N$

$$(\,(\,b_1, \ldots, b_N\,), Q^N(\,(\,b_1, \ldots, b_N\,)\,)\,) \prec^{Q^N}{}_{[S_1, \ldots, S_N]/[s_1, \ldots, s_N]} (\,(\,c_1, \ldots, c_N\,), Q^N(\,(\,c_1, \ldots, c_N\,)\,)\,)$$

$$\text{iff}$$

$$(\,b_1, \ldots, b_N\,) <_{[S_1, \ldots, S_N]/[s_1, \ldots, s_N]} (\,c_1, \ldots, c_N\,)$$

---

## Multi-Array: Cartesian Extension.

**Definition 3.2 :** Let $Q^N = TI^{[1/N]}{}_{[S_1, \ldots, S_N]/[s_1, \ldots, s_N]}$ be an array. Let $\mathcal{A}^{[1/N]}{}_{[S_1, \ldots, S_N]/[s_1, \ldots, s_N]}$, be indexing hierarchy. We define array $Q^N$ depth **0** ()-Cartesian Projection, $P^N{}_{[]}$, as follows :

$$Q^N{}_{[]} \equiv \{\,(\,(\,a_1, \ldots, a_N\,), Q^N(\,(\,a_1, \ldots, a_N\,)\,)\,):$$

$$(\,a_1, \ldots, a_N\,) \in I^{[1/N]}{}_{[S_1, \ldots, S_N]/[s_1, \ldots, s_N]}\,\}$$

---

**Definition 3.3 :** Let $Q^N = TI^{[1/N]}{}_{[S_1, \ldots, S_N]/[s_1, \ldots, s_N]}$ be an array. Let $0 < M < N$. Let $(a_1, \ldots, a_M) \in I^{[1/M]}{}_{[S_1, \ldots, S_M]/[s_1, \ldots, s_M]}$. We define array $Q^N$ depth $M$ $(a_1, \ldots, a_M)$-Cartesian Projection, $Q^{N-M}{}_{[a_1, \ldots, a_M]}$, as follows:

$$Q^{N-M}{}_{[a_1, \ldots, a_M]} \equiv$$

$$\{\,(\,(\,a_1, \ldots, a_M, a_{M+1}, \ldots, a_N\,), Q^N(\,(\,a_1, \ldots, a_M, a_{M+1}, \ldots, a_N\,)\,)\,):$$

$$(\,a_{M+1}, \ldots, a_N\,) \in I^{[M+1/N-M]}{}_{[S_{M+1}, \ldots, S_N]/[s_{M+1}, \ldots, s_N]}\,\} \equiv$$

$$\{\,(\,(\,c_1, \ldots, c_N\,), Q^N(\,(\,c_1, \ldots, c_N\,)\,)\,):$$

$$(\,c_1, \ldots, c_N\,) \in \{a_1\} \times \ldots \times \{a_M\} \times I^{[M+1/N-M]}{}_{[S_{M+1}, \ldots, S_N]/[s_{M+1}, \ldots, s_N]}\,\}$$

---

**Definition 3. 4 :** Let $Q^N = TI^{[1/N]}{}_{[S_1, \ldots, S_N]/[s_1, \ldots, s_N]}$ be an array. Let $I^{[1/N]}{}_{[S_1, \ldots, S_N]/[s_1, \ldots, s_N]}$ be array $Q^N$ indexing set. Let $(a_1, \ldots, a_N) \in I^{[1/N]}{}_{[S_1, \ldots, S_N]/[s_1, \ldots, s_N]}$.

We define array $Q^N$ depth $N$ $(a_1, \ldots, a_N)$-Cartesian Projection, $Q^0{}_{[a_1, \ldots, a_N]}$, as follows :

$$P^0{}_{[a_1, \ldots, a_N]} \equiv \{\,(\,(\,a_1, \ldots, a_N\,), Q^N(\,(\,a_1, \ldots, a_N\,)\,)\,)\,\}.$$

---

**Definition 3. 5 :** Let $Q^N = TI^{[1/N]}{}_{[S_1, \ldots, S_N]/[s_1, \ldots, s_N]}$ be an array. We define multi-array $Q^N$ Cartesian

Extension, $\mathcal{A}_{Q^N}$, as an inclusion-ordered totality of array $Q^N$ level **0** through level **N** Cartesian Projections.

---

**Lemma 3.1:** Let $Q^N = TI^{[1/N]}_{[S_1, ..., S_N]/[s_1, ..., s_N]}$ be an array. Let $0 < M < N$.

Let $I^{[1/M]}_{[S_1, ..., S_M]/[s_1, ..., s_M]}$ be array $Q^N$ partial indexing set.

Let $(a_1, ..., a_M) \in I^{[1/M]}_{[S_1, ..., S_M]/[s_1, ..., s_M]}$.

Let $Q^{N-M}_{[a_1, ..., a_M]}$ be array $Q^N$ $(a_1, ..., a_M)$-Cartesian Projection.

Then array $Q^{N-M}_{[a_1, ..., a_M]}$ is an $S_{M+1} * ... * S_N$ long, $\prec^{Q^N}_{[S_1, ..., S_N]/[s_1, ..., s_N]}$-*contiguous* subinterval of interval $Q^N$.

**Proof:**

By definition, $Q^{N-M}_{[a_1, ..., a_M]} \equiv \{ ((c_1, ..., c_N), Q^N(c_1, ..., c_N))$,

$(c_1, ..., c_N) \in \{a_1\} \times ... \times \{a_M\} \times I^{[M+1/N-M]}_{[S_{M+1}, ..., S_N]/[s_{M+1}, ..., s_N]} \}$ is an array.

By definition, $Q^{N-M}_{[a_1, ..., a_M]}$ is $S_{M+1} * ... * S_N$ sized array.

We next prove that array $Q^{N-M}_{[a_1, ..., a_M]}$ is an $\prec^{Q^N}_{[S_1, ..., S_N]/[s_1, ..., s_N]}$-contiguous subinterval of interval $Q^N$.

Array's $Q^{N-M}_{[a_1, ..., a_M]}$ $\prec^{Q^N}_{[S_1, ..., S_N]/[s_1, ..., s_N]}$-first element is pair

$A = ((a_1, ..., a_M, s_{M+1} + 1, ..., s_N + 1), Q^N((a_1, ..., a_M, s_{M+1} + 1, ..., s_N + 1)))$,

Array's $Q^{N-M}_{[a_1, ..., a_M]}(Q^{N-M}_{[a_1, ..., a_M]})$ $\prec^{Q^N}_{[S_1, ..., S_N]/[s_1, ..., s_N]}$-last element is pair

$B = ((a_1, ..., a_M, s_{M+1} + S_{M+1}, ..., s_N + S_N),$

$Q^N((a_1, ..., a_M, s_{M+1} + S_{M+1}, ..., s_N + S_N)))$.

Let's assume that array $Q^{N-M}_{[a_1, ..., a_M]}$ is not an $\prec^{Q^N}_{[S_1, ..., S_N]/[s_1, ..., s_N]}$-contiguous subinterval of array $Q^N$.

Let $C \in Q^N$ be such that

(a) $A \prec^{Q^N}_{[S_1, ..., S_N]/[s_1, ..., s_N]} C \prec^{Q^N}_{[S_1, ..., S_N]/[s_1, ..., s_N]} B$, and

(b) $C$ is not an element of array $E^{N-M}_{[a_1, ..., a_M]}$

Let $(c_1, ..., c_N) = (Q^N)^{-1}(C)$.

By $\prec^{Q^N}_{[S_1, ..., S_N] / [s_1, ..., s_N]}$ definition,

inequality $C \prec^{Q^N}_{[S_1, ..., S_N] / [s_1, ..., s_N]} B$ implies that

$$(c_1, ..., c_N) \prec_{[S_1, ..., S_N] / [s_1, ..., s_N]} (a_1, ..., a_M, s_{M+1} + S_{M+1}, ..., s_N + S_N)$$

That, in turn, implies that

$$(c_1, ..., c_M) \leq_{[S_1, ..., S_M] / [s_1, ..., s_M]} (a_1, ..., a_M)$$

By $\prec^{Q^N}_{[S_1, ..., S_N] / [s_1, ..., s_N]}$ definition,

inequality $A \prec^{Q^N}_{[S_1, ..., S_N] / [s_1, ..., s_N]} C$ implies that

$$(a_1, ..., a_M, s_{M+1} + 1, ..., s_N + 1) \prec_{[S_1, ..., S_N] / [s_1, ..., s_N]} (c_1, ..., c_N)$$

That, in turn, implies that

$$(a_1, ..., a_M) \leq_{[S_1, ..., S_M] / [s_1, ..., s_M]} (c_1, ..., c_M)$$

Thus

$$(c_1, ..., c_M) = (a_1, ..., a_M)$$

That in turn means that $C \in Q^{N-M}_{[a_1, ..., a_M]}$.

**Q.E.D.**

∎

**Lemma 3.2:** Let $Q^N = TI^{[1/N]}_{[S_1, ..., S_N] / [s_1, ..., s_N]}$ be an array.

Let **L, M** be natural numbers such that $0 \leq L \leq N$ and $0 \leq M \leq N$.

Let $I^{[1/L]}_{[S_1, ..., S_L] / [s_1, ..., s_L]}$, $I^{[1/M]}_{[S_1, ..., S_N] / [s_1, ..., s_M]}$ be array $Q^N$ partial indexing sets.

Let $(a_1, ..., a_L) \in I^{[1/L]}_{[S_1, ..., S_L] / [s_1, ..., s_L]}$ and $(b_1, ..., b_M) \in I^{[1/M]}_{[S_1, ..., S_M] / [s_1, ..., s_M]}$

Let $Q^{N-L}_{[a_1, ..., a_L]}$ and $Q^{N-M}_{[b_1, ..., b_M]}$ be array $Q^N$ depth **L** and depth **M** Cartesian Projections respectively.

Let array $Q^{N-L}_{[a_1, ..., a_L]}$ be a subset of multi-array $Q^{N-M}_{[b_1, ..., b_M]}$.

Then **L > M.**

**Proof:**

By **theorem 3 / 1**, array $Q^{N-L}_{[a_1, ..., a_L]}$ size is $S_{L+1} * ... * S_N$, and array $Q^{N-M}_{[a_1, ..., a_M]}$ size is

$S_{M+1} * ... * S_N.$

Since array $Q^{N-L}_{[a_1, ..., a_L]}$ is a subset of array $Q^{N-M}_{[b_1, ..., b_M]}$

that means that $S_{L+1} * ... * S_N < S_{M+1} * ... * S_N.$

Thus $L > M.$

Q.E.D.

∎

**Lemma 3.3:** Let $Q^N = TI^{[1/N]}_{[S_1, ..., S_N]/[s_1, ..., s_N]}$ be an array. Let $M$ be a natural number such that $0 < M \leq N$. Let $I^{[1/M]}_{[S_1, ..., S_N]/[s_1, ..., s_M]}$ be array $Q^N$ partial indexing set.
Let $(a_1, ..., a_M), (b_1, ..., b_M) \in I^{[1/M]}_{[S_1, ..., S_M]/[s_1, ..., s_M]}.$
Let $Q^{N-M}_{[a_1, ..., a_M]}$ and $Q^{N-M}_{[a_1, ..., a_M]}$ be array $Q^N$ depth $M$ Cartesian Projections.
Then $(a_1, ..., a_M) \neq (b_1, ..., b_M)$

iff

sets $Q^{N-M}_{[a_1, ..., a_M]}$ and $Q^{N-M}_{[b_1, ..., b_M]}$ are disjoint.

**Proof:**

Multi-array $Q^{N-M}_{[a_1, ..., a_M]}$ and multi-array $Q^{N-M}_{[a_1, ..., a_M]}$ are subarrays of multi-array $Q^N$.

Multi-array $Q^{N-M}_{[a_1, ..., a_M]}$ indexing set is

$$\{a_1\} \times ... \times \{a_M\} \times I^{[M+1/N-M]}_{[S_{M+1}, ..., S_N]/[s_{M+1}, ..., s_N]}.$$

Multi-array $Q^{N-M}_{[b_1, ..., b_M]}$ indexing set is

$$\{b_1\} \times ... \times \{b_M\} \times I^{[M+1/N-M]}_{[S_{M+1}, ..., S_N]/[s_{M+1}, ..., s_N]}$$

Q.E.D.

∎

**Theorem 3.1:** Let $Q^N = TI^{[1/N]}_{[S_1, ..., S_N]/[s_1, ..., s_N]}$ be an array. Let $L, M$ be natural numbers such that $1 \leq L \leq N$ and $1 \leq M \leq N$.

Let $I^{[1/L]}_{[S_1, ..., S_L]/[s_1, ..., s_L]}$ and $I^{[1/M]}_{[S_1, ..., S_N]/[s_1, ..., s_M]}$ be array $Q^N$ partial indexing sets.

Let $(a_1, ..., a_L) \in I^{[1/L]}_{[S_1, ..., S_L]/[s_1, ..., s_L]}.$
Let $(b_1, ..., b_M) \in I^{[1/M]}_{[S_1, ..., S_M]/[s_1, ..., s_M]}$

Let $\mathbf{Q^{N-L}}_{[a_1, ..., a_L]}$ and $\mathbf{Q^{N-M}}_{[b_1, ..., b_M]}$ be array $\mathbf{Q^N}$ Cartesian Projections of depth $\mathbf{M}$ and depth $\mathbf{L}$ respectively.

Let $\mathbf{Q^{N-M}}_{[b_1, ..., b_M]}$ and $\mathbf{Q^{N-L}}_{[a_1, ..., a_L]}$ be array $\mathbf{Q^N}$ Cartesian Projections of depth $\mathbf{M}$ and depth $\mathbf{L}$ respectively.

Then

multi-array $\mathbf{Q^{N-L}}_{[a_1, ..., a_L]}$ is a subset of multi-array $\mathbf{Q^{N-M}}_{[b_1, ..., b_M]}$

<div align="center">iff</div>

(a) $\mathbf{L > M}$, and

(b) $(\mathbf{a_1}, ..., \mathbf{a_M}) = (\mathbf{b_1}, ..., \mathbf{b_M})$

**Proof:**

Let's assume that conditions **(a)** and **(b)** hold.

Then,

multi-array's $\mathbf{Q^{N-M}}_{[b_1, ..., b_M]}$ indexing set is

$$\{\mathbf{b_1}\} \times ... \times \{\mathbf{b_M}\} \times \mathbf{I}^{[M+1/N-M]}_{[S_{M+1}, ..., S_N] / [s_{M+1}, ..., s_N]}, \text{ and}$$

multi-array's $\mathbf{Q^{N-L}}_{[a_1, ..., a_L]}$ indexing set is

$$\{\mathbf{a_1}\} \times ... \times \{\mathbf{a_L}\} \times \times \mathbf{I}^{[L+1/N-L]}_{[S_{L+1}, ..., S_N] / [s_{L+1}, ..., s_N]} =$$

$$\{\mathbf{b_1}\} \times ... \times \{\mathbf{b_M}\} \times \{\mathbf{a_{M+1}}\} \times ... \times \{\mathbf{a_L}\} \times \mathbf{I}^{[L+1/N-L]}_{[S_{L+1}, ..., S_N] / [s_{L+1}, ..., s_N]}.$$

By definition, since $\mathbf{M < N}$, multi-array $\mathbf{Q^{N-L}}_{[a_1, ..., a_L]}$ is a subset of multi-array $\mathbf{P^{N-M}}_{[a_1, ..., a_M]}$.

Thus, since it is assumed that $(\mathbf{a_1}, ..., \mathbf{a_M}) = (\mathbf{b_1}, ..., \mathbf{b_M})$, multi-array $\mathbf{Q^{N-L}}_{[a_1, ..., a_L]}$ is a subset of multi-array $\mathbf{Q^{N-M}}_{[b_1, ..., b_M]}$.

Let's assume that multi-array $\mathbf{Q^{N-L}}_{[a_1, ..., a_L]}$ is a subset of multi-array $\mathbf{Q^{N-M}}_{[b_1, ..., b_M]}$.

We show next that conditions **(a)** and **(b)** hold.

By **lemma 3.2**, since multi-array $\mathbf{Q^{N-L}}_{[a_1, ..., a_L]}$ is a subset of multi-array $\mathbf{P^{N-M}}_{[b_1, ..., b_M]}$,

**L > M** holds.

Then, since **L > M,** multi-array $Q^{N-L}{}_{[a_1, ..., a_L]}$ is a subset of multi-array $Q^{N-M}{}_{[a_1, ..., a_M]}$.

Since, by assumption, multi-array $Q^{N-L}{}_{[a_1, ..., a_L]}$ is a subset of multi-array $Q^{N-M}{}_{[b_1, ...-Z, b_M]}$, that means that multi-array $Q^{N-L}{}_{[a_1, ..., a_L]}$ is a subset of $Q^{N-M}{}_{[a_1, ..., a_M]} \cap Q^{N-M}{}_{[b_1, ..., b_M]}$ :

multi-arrays $Q^{N-M}{}_{[a_1, ..., a_M]}$ and $Q^{N-M}{}_{[b_1, ..., b_M]}$ are not disjoint.

By lemma **3 / 3,** $(a_1, ..., a_M) = (b_1, ..., b_M)$.

**Q.E.D.**

■

**Theorem 3.2:** Let $Q^N = TI^{[1/N]}{}_{[S_1, ..., S_N] / [s_1, ..., s_N]}$ be an array. Let $\mathcal{A}_{Q^N}$ be multi-array $Q^N$ Cartesian Extension.

Then $\mathcal{A}_{Q^N}$ is an $[S_1, ..., S_N] / [s_1, ..., s_N]$ indexing order hierarchy.

**Proof:**

Let $\mathcal{A}^{[1/N]}{}_{[S_1, ..., S_N] / [s_1, ..., s_N]}$ be an $[S_1, ..., S_N] / [s_1, ..., s_N]$ bi ndexing hierarchy.

Let map $T_{Q^N} : \mathcal{A}^{[1/N]}{}_{[S_1, ..., S_N] / [s_1, ..., s_N]} \to \mathcal{A}_{Q^N}$ be such that

for $(c_1, ..., c_M) \in \mathcal{A}^{[1/N]}{}_{[S_1, ..., S_N] / [s_1, ..., s_N]}$  $T_{Q^N}((c_1, ..., c_M)) = Q^{N-M}{}_{[c_1, ..., c_M]}$.

Map $T_{Q^N}$ is an onto map. By **theorem 3 / 1**, map $T_{Q^N}$ is an onto, order-preserving map.

**Q.E.D.**

■

**Theorem 3.3:** Let $Q^N$ be an $[S_1, ..., S_N] / [s_1, ..., s_N]$ multi-array.

Multi-array $Q^N$ Cartesian Extensionis an $[S_1, ..., S_N] / [s_1, ..., s_N]$ multi-cube.

**Proof:**

Let $\mathcal{A}_{Q^N}$ be multi-array $Q^N$ Cartesian Extension.

By **theorem 3.2**, $\mathcal{A}_{Q^N}$ is an $[S_1, ..., S_N] / [s_1, ..., s_N]$ indexing order hierarchy.

By **theorem 2.5**, $\mathcal{A}_{Q^N}$ is an $[S_1, ..., S_N] / [s_1, ..., s_N]$ multi-cube.

**Q.E.D.**

∎

**Lemma 3.4:** Let $Q^N = TI^{[1/N]}_{[S_1, ..., S_N] / [s_1, ..., s_N]}$ be an array. Let $\mathcal{A}_{Q^N}$ be array $Q^N$ Cartesian extension. Let **L** be natural number such that $0 < L \leq N$.

Let $(a_1, ..., a_L) \in I^{[1/L]}_{[S_1, ..., S_L] / [s_1, ..., s_L]}$.

Let $Q^{N-L}_{[a_1, ..., a_L]} \in \mathcal{A}_{Q^N}$ be array $Q^N$ depth **L** Cartesian Projection.

Then $Q^{N-L}_{[a_1, ..., a_L]}$ is hierarchy $\mathcal{A}_{Q^N}$ level **L** node.

**Proof:**

By **theorem 3.2**, hierarchy $\mathcal{A}_{Q^N}$ is an $[S_1, ..., S_N] / [s_1, ..., s_N]$ indexing order hierarchy.

Let $\mathcal{A}^{[1/N]}_{[S_1, ..., S_N] / [s_1, ..., s_N]}$ be $[S_1, ..., S_N] / [s_1, ..., s_N]$ indexing hierarchy.

Let $T : \mathcal{A}^{[1/N]}_{[S_1, ..., S_N] / [s_1, ..., s_N]} \to \mathcal{A}_{Q^N}$ such that

for $(a_1, ..., a_L) \in \mathcal{A}^{[1/N]}_{[S_1, ..., S_N] / [s_1, ..., s_N]}$

$T( (a_1, ..., a_L) ) = Q^{N-L}_{[a_1, ..., a_L]}$.

By **theorem 3.2**, map **T** is an onto, one-to-one, order-preserving map.

By **theorem 1.7**, node $(a_1, ..., a_L)$ is hierarchy $\mathcal{A}^{[1/N]}_{[S_1, ..., S_N] / [s_1, ..., s_N]}$ level **L** node.

By **theorem 1.9**, node $T( (a_1, ..., a_L) ) = Q^{N-L}_{[a_1, ..., a_L]}$ is hierarchy $\mathcal{A}_{Q^N}$ level **L** node.

**Q.E.D.**

∎

**Theorem 3.4 :** Let $Q^N = TI^{[1/N]}_{[S_1, ..., S_N]/[s_1, ..., s_N]}$ be an array.

Let $\mathcal{A}_{Q^N}$ be array $Q^N$ Cartesian extension. Let $L$ be natural number such that $0 \leq L < N$.

Let $(a_1, ..., a_L) \in \mathcal{A}^{[1/N]}_{[S_1, ..., S_N]/[s_1, ..., s_N]}$

Let $Q^{N-L}_{[a_1, ..., a_L]}$ be hierarchy $\mathcal{A}_{Q^N}$ level $L$ node.

Then node $Q^{N-L}_{[a_1, ..., a_L]}$ has $S_{L+1} [S_{L+1}] / [s_{L+1}]$-indexed set of children

$$\{ Q^{N-L-1}_{[a_1, ..., a_L, a_{L+1}]} : a_{L+1} \in I^{L+1}_{S_{L+1}/s_{L+1}} \}$$

**Proof:**

Let map $T : \mathcal{A}^{[1/N]}_{[S_1, ..., S_N]/[s_1, ..., s_N]} \to \mathcal{A}_{Q^N}$ be such that

for $(c_1, ..., c_M) \in \mathcal{A}^{[1/N]}_{[S_1, ..., S_N]/[s_1, ..., s_N]}$

$T((c_1, ..., c_M)) = Q^{N-M}_{[c_1, ..., c_M]}$.

Then $T((a_1, ..., a_L)) = Q^{N-L}_{[a_1, ..., a_L]}$.

By **theorem 1.9**, set $\{ (a_1, ..., a_L, a_{L+1}) : a_{L+1} \in I^{L+1}_{S_{L+1}/s_{L+1}} \}$ is hierarchy's

$\mathcal{A}^{[1/N]}_{[S_1, ..., S_N]/[s_1, ..., s_N]}$ $[S_{L+1}]/[s_{L+1}]$-indexed set of node's $(a_1, ..., a_L)$ children.

By **lemma 3.4** and **theorem 1/10,** set $\{ Q^{N-L-1}_{[a_1, ..., a_L, a_{L+1}]} : a_{L+1} \in I^{L+1}_{S_{L+1}/s_{L+1}} \}$ is

hierarchy's $\mathcal{A}^{[1/N]}_{[S_1, ..., S_N]/[s_1, ..., s_N]}$ $[S_{L+1}]/[s_{L+1}]$-indexed set of node $Q^{N-L}_{[a_1, ..., a_L]}$ children.

   **Q.E.D.**
∎

**Lemma 3.5:** Let $Q^N = TI^{[1/N]}_{[S_1, ..., S_N]/[s_1, ..., s_N]}$ be an array. Let hierarchy $\mathcal{A}_{Q^N}$ be array $Q^N$

Cartesian extension. Let $(a_1, ..., a_N) \in I^{[1/N]}_{[S_1, ..., S_N]/[s_1, ..., s_N]}$.

Let $Q^0_{[a_1, ..., a_N]} = \{ ((a_1, ..., a_N), Q^N((a_1, ..., a_N))) \}$ be multi-array $Q^N$ depth $N$

Cartesian Projection.

Then node $Q^0_{[a_1, ..., a_N]}$ is hierarchy's $\mathcal{A}_{Q^N}$ data node.

**Proof:**

By **lemma 3.4,** node $Q^0_{[a_1, ..., a_N]}$ is hierarchy's $\mathcal{A}_{Q^N}$ level **N** node.

By **theorem 3.3,** hierarchy $\mathcal{A}_{Q^N}$ is an $[S_1, ..., S_N] / [s_1, ..., s_N]$ multi-cube.

By **theorem 2.2,** node $Q^0_{[a_1, ..., a_N]}$ is hierarchy's $\mathcal{A}_{Q^N}$ data-node.

Q.E.D.
■

**Theorem 3.5 :** Let $Q^N = TI^{[1/N]}_{[S_1, ..., S_N] / [s_1, ..., s_N]}$ be an array. Let hierarchy $\mathcal{A}_{Q^N}$ be array $Q^N$ Cartesian extension. Then set

$\{ \, \{ \, ( \, (a_1, ..., a_N) \, ), Q^N ( \, ( a_1, ..., a_N ) \, ) \, ) \, \} : (a_1, ..., a_N) \in I^{[1/N]}_{[S_1, ..., S_N] / [s_1, ..., s_N]} \, \}$ is

hierarchy $\mathcal{A}_{Q^N}$ data-set.

**Proof:**

Follows directly from **Lemma 3.5.**

Q.E.D.
■

## Type-* Multi-Array.

**Definition 3. 9 :** Let **P** be a type. We define multi-array $Q^N$ as a multi-array of type **P** if multi-array $Q^N$ range's elements are of type **P**.

# IV. Quantizing a Multi-Cube.

**Definition 4.1:** We define quantizing function as a function that maps linearly ordered sets of scalars to scalars.

**Definition 4.2:** Let **M** be a positive natural number. We define quantizing function of order **[M]** as a quantizing function that maps **[M]**-indexed sets of scalars to scalars.

**Definition 4.3:** Let **M** be a natural number. Let **m** be an integer number. We define quantizing function of order **[M] / [m]** as a quantizing function that maps **[M] / [m]**-indexed sets of scalars to scalars.

---

**Definition 4.4:** Let **P** be a type. We define quantizing function of type **P** as a function that maps linearly ordered sets of type **P** scalars to type **P** scalars.

---

**Definition 4.5:** Let **P** and **Q** be types. Let $_P T_Q$ be type **P** to type **Q** converter.
Let $C^N$ be $[S_1, ..., S_N] / [s_1, ..., s_N]$ multi-cube of type **P**. Let $f_1, ..., f_N$ be type-**Q** quantizing functions of order $[S_1] / [s_1], ..., [S_N] / [s_N]$ respectively. Let $_P T_Q$ be type **P** to type **Q** converter.

We define quantizing type-**P** multi-cube $C^N$, in terms of type-**Q** quantizing functions $f_1, ..., f_N$, and in terms of $_P T_Q$ type converter, as mapping multi-cube $C^N$ to a type-**Q** value by:

   (a) quantizing each of multi-cube $C^{N-1}_{[a_i]}$, $a_i \in I^1_{[S_1]/[s_1]}$, depth **1** parsing ranges in terms of quantizing functions $f_2 ... f_{N-1}$, and in terms of $_P T_Q$ type converter, thus generating an $[S_1] / [s_1]$-indexed type-**Q** set **F** of type-**Q** values.

   (b) Mapping multi-cube $C^N$ to $f_1(F)$ type-**Q** return value.

---

**Definition 4.6:** Let **P** and **Q** be types. Let $_P T_Q$ be type **P** to type **Q** converter. Let $C^1$ be $[S] / [s]$ multi-cube of type **P**. Let **f** be type-**Q** quantizing function of order $[S] / [s]$.

Let $\mathcal{D}^1$ be multi-cube $C^1$ $[S] / [s]$-indexed data-set.

We define quantizing multi-cube $C^1$, in terms of type-**P** quantifying function $f_1$ of order $[S / s]$, and in terms of $_P T_Q$ type converter, as mapping multi-cube $C^1$ to $f_1 ( _P T_Q ( \mathcal{D}^1 ) )$ return value.

---

**Definition 4.7:** Let $\mathcal{A}$ and $\mathcal{B}$ be $[S_1, ..., S_N] / [s_1, ..., s_N]$ multi-cubes.
We define multi-cubes $\mathcal{A}$ and $\mathcal{B}$ as equivalent, $\mathcal{A} \approx \mathcal{B}$, if multi-cubes $\mathcal{A}$ and $\mathcal{B}$ share a data-set.

---

**Lemma 4.1 :** Let $\mathcal{X}^N$ and $\mathcal{Y}^N$ be $[S_1, ..., S_N] / [s_1, ..., s_N]$ multi-cubes. Let $\mathcal{X}^N \approx \mathcal{Y}^N$. Let $I^{[1/N]}_{[S_1, ..., S_N]/[s_1, ..., s_N]}$ their shared indexing set. Let $a \in I^1_{[S_1]/[s_1]}$. Let $[S_2, ..., S_N] / [s_2, ..., s_N]$ multi-cubes $\mathcal{X}^{N-1}_{[a]}$ and $\mathcal{Y}^{N-1}_{[a]}$ be multi-cubes' $\mathcal{X}^N$ and $\mathcal{Y}^N$ depth **1** (a)-parsing ranges respectively.

Then $\mathcal{X}^{N-1}_{[a]} \approx \mathcal{Y}^{N-1}_{[a]}$.

**Proof :**

By theorem 2 / 7, node $[a_2, ..., a_N] \in \mathcal{X}^{N-1}_{[a]}$ is multi-cube $\mathcal{X}^{N-1}_{[a]}$ terminal node.

By definition, node $[a_2, ..., a_N] \in \mathcal{X}^{N-1}_{[a]}$ is node $[a, a_2, ..., a_N] \in \mathcal{X}^{N}$.

Since $[a, a_2, ..., a_N]$, by virtue of being multi-cube $\mathcal{X}^N$ level **N** node, is multi-cube $\mathcal{X}^N$ terminal node, and multi-cubes $\mathcal{X}^N$ and $\mathcal{Y}^N$ share their data-sets, $[a, a_2, ..., a_N]$ is multi-cube $\mathcal{Y}^N$ terminal node as well.

By definition, node node $[a, a_2, ..., a_N] \in \mathcal{Y}^N$ is node $[a_2, ..., a_N] \in \mathcal{Y}^{N-1}_{[a]}$.

By theorem 2 / 7, node $[a_2, ..., a_N]$ is multi-cube $\mathcal{Y}^{N-1}_{[a]}$ terminal node as well.

Thus, if $\mathcal{D}^{N-1}_{[a]}$ is multi-cube $\mathcal{Y}^{N-1}_{[a]}$ data-set as well.

**Q.E.D.**
∎

**Theorem 4.1 :** Let **P, Q** be types. Let $_P T_Q$ be type **P** to type **Q** converter. Let $\mathcal{X}^N$ and $\mathcal{Y}^N$ be multi-cubes of type **P**. Let $\mathcal{X}^N \approx \mathcal{Y}^N$.

Then, quantizing either multi-cube $\mathcal{C}^N$ or multi-cube $\mathcal{D}^N$, in terms of type-**Q** quantizing functions $f_1$, ..., $f_N$ of order $[S_1] / [s_1]$, ..., $[S_N] / [s_N]$ respectively, and in terms of $_P T_Q$ type converter, generates identical result.

**Proof:**

The proof is by dimensional induction.

Let $\mathcal{X}^1$ and $\mathcal{Y}^1$ be $[S] / [s]$ multi-cubes that share $[S] / [s]$ data-array $D^1$ of type **P**.

That means that $\mathcal{X}^1$ and $\mathcal{Y}^1$ share type **P** $[S] / [s]$-indexed set $\mathcal{D}^1$ as their data-set.

By definition, quantizing either, in terms of type **Q** quantizing function $f_1$ of order $[S] / [s]$, means mapping each of multi-cubes $\mathcal{X}^1$ and $\mathcal{Y}^1$ to function $f_1(_P T_Q(\mathcal{D}^1))$ return value.

Let $\mathcal{X}^N$ and $\mathcal{Y}^N$ be $[S_1, ..., S_N] / [s_1, ..., s_N]$ multi-cubes. Let $I^{[1/N]}_{[S_1, ..., S_N] / [s_1, ..., s_N]}$ be multi-cubes $\mathcal{X}^N$ and $\mathcal{Y}^N$ shared indexing set.

Let $a_i \in I^1_{[S_1]/[s_1]}$, $i = 1, ..., S_1$

Then, by **Lemma 4 / 1,** for multi-cubes $\mathcal{X}^N$ and $\mathcal{Y}^N$ depth **1** parsing ranges $\mathcal{X}^{N-1}_{[\,a_i\,]}$ and $\mathcal{Y}^{N-1}_{[\,a_i\,]}$, it holds that $\mathcal{X}^{N-1}_{[\,a_i\,]} \approx \mathcal{Y}^{N-1}_{[\,a_i\,]}$, $a_i \in \mathbf{I}^1_{[S_1]/[s_1]}$.

Let's assume that quantizing equivalent **N - 1**-dimensional multi-cubes, in terms of a shared set of quantizing functions, and in terms of a shared type converter, yields identical results.

By the assumption, $[S_1] / [s_1]$-indexed type **Q** set $F_1$ that is generated as a result of quantizing each of multi-cube's $\mathcal{X}^N$ $\mathcal{X}^{N-1}_{[\,a_i\,]}$ ($a_i \in \mathbf{I}^1_{[S_1]/[s_1]}$) parsing ranges, in terms of type **Q** quantizing functions $f_2 \ldots f_N$ of order $[S_2] / [s_2]$, ..., $[S_N] / [s_N]$ respectively, and in terms of $_P T_Q$ type converter, is identical to $[S_1] / [s_1]$-indexed type **Q** set $F_2$ generated as a result of quantizing each of multi-cube $\mathcal{Y}^N$ $\mathcal{Y}^{N-1}_{[\,a\,]}$ parsing ranges, $a \in \mathbf{I}^1_{[S_1]/[s_1]}$, in terms of type **Q** quantizing functions $f_2 \ldots f_N$ of order $[S_2] / [s_2]$, ..., $[S_N] / [s_N]$ respectively, and in terms of $_P T_Q$ type converter.

Thus quantizing either multi-cube $\mathcal{X}^N$ or multi-cube $\mathcal{Y}^N$ in terms of type **Q** quantizing functions $f_1 \ldots f_N$ of order $[S_1]/[s_1]$, ..., $[S_N]/[s_N]$ respectively, and in terms of $_P T_Q$ type converter, consists of mapping each of the multi-cubes to $f_1(F_1)$ return value.

**Q.E.D.**

# Quantizing Multi-Array : Definitions

## Quantizing Multi-Array Globally : Definitions

**Definition 5.1 :** Let multi-array $\mathbf{Q}^N = \mathbf{TI}^{[1/N]}_{[S_1, \ldots, S_N]/[s_1, \ldots, s_N]}$ be of type **P**. Let $f_1, \ldots, f_N$ be type-**Q** quantizing functions of order $[S_1] / [s_1], \ldots, [S_N] / [s_N]$ respectively. Let $_P T_Q$ be type **P** to type **Q** converter.

We define quantizing type **P.** multi-array $\mathbf{Q}^N$ globally, in terms of type **Q** quantizing functions $f_1, \ldots, f_N$ of order $[S_1]/[s_1] \ldots [S_N]/[s_N]$ respectively, and in terms of $_P T_Q$ type converter, as quantizing multi-array's $\mathbf{Q}^N$ Cartesian Extension $\mathcal{A}_{\mathbf{Q}^N}$ in terms of quantizing functions $f_1, \ldots, f_N$ of order $[S_1]/[s_1] \ldots [S_N]/[s_N]$ respectively, and in terms of $_P T_Q$ type converter.

# Quantizing Multi-Array Locally : Definitons.

**Lemma 5.1:** Let $\mathcal{A}^{[1/N]}_{[S_1, ..., S_N]/[s_1, ..., s_N]}$ and $\mathcal{A}^{[1/N]}_{[T_1, ..., T_N]/[t_1, ..., t_N]}$ be indexing hierarchies. Let $T^{[1/N]}_{[S_1, ..., S_N]/[s_1, ..., s_N]}$ be hierarchy's $\mathcal{A}^{[1/N]}_{[S_1, ..., S_N]/[s_1, ..., s_N]}$ indexing set. Let $T^{[1/N]}_{[T_1, ..., T_N]/[t_1, ..., t_N]}$ be hierarchy's $\mathcal{A}^{[1/N]}_{[T_1, ..., T_N]/[t_1, ..., t_N]}$ indexing set.

Then

hierarchy $\mathcal{A}^{[1/N]}_{[T_1, ..., T_N]/[t_1, ..., t_N]}$ is a subhierarchy of hierarchy $\mathcal{A}^{[1/N]}_{[S_1, ..., S_N]/[s_1, ..., s_N]}$
**iff**
indexing set $T^{[1/N]}_{[T_1, ..., T_N]/[t_1, ..., t_N]}$ is a subset of indexing set $T^{[1/N]}_{[S_1, ..., S_N]/[s_1, ..., s_N]}$.

**Proof :**

Let $I^{[1/N]}_{[S_1, ..., S_N]/[s_1, ..., s_N]}$ be hierarchy $\mathcal{A}^{[1/N]}_{[S_1, ..., S_N]/[s_1, ..., s_N]}$ indexing set.
By definition, set $\mathcal{A}^{[1/N]}_{[T_1, ..., T_N]/[t_1, ..., t_N]}$ is a union

$\{\ (\ )\ \} \cup \{\ I^{[1/1]}_{[S_1]/[s_1]}\} \cup \{\ I^{[1/2]}_{[S_1, S_2]/[s_1, s_2]}\} \cup ... \cup \{\ I^{[1/N]}_{[S_1, ..., S_N]/[s_1, ..., s_N]}\ \}$

Let $I^{[1/N]}_{[T_1, ..., T_N]/[t_1, ..., t_N]}$ be hierarchy $\mathcal{A}^{[1/N]}_{[T_1, ..., T_N]/[t_1, ..., t_N]}$ indexing set.
By definition, set $\mathcal{A}^{[1/N]}_{[T_1, ..., T_N]/[t_1, ..., t_N]}$ is a union

$\{\ (\ )\ \} \cup \{\ I^{[1/1]}_{[T_1]/[t_1]}\} \cup \{\ I^{[1/2]}_{[T_1, T_2]/[t_1, t_2]}\} \cup ... \cup \{\ I^{[1/N]}_{[T_1, ..., T_N]/[t_1, ..., t_N]}\ \}$

Indexing set $T^{[1/N]}_{[T_1, ..., T_N]/[t_1, ..., t_N]}$ is a subset of indexing set $T^{[1/N]}_{[S_1, ..., S_N]/[s_1, ..., s_N]}$

**iff**

indexing sets $T^{i}_{T_i/t_i}$ are subsets of indexing sets $T^{i}_{S_i/s_i}$ respectively, $i = 1, ..., N$

**iff**

indexing sets $T^{[1/i]}_{[T_1, ..., T_i]/[t_1, ..., t_i]}$ are subsets of indexing sets $T^{[1/i]}_{[S_1, ..., S_i]/[s_1, ..., s_i]}$ respectively, $i = 1, ..., N$

**iff**

set $\mathcal{A}^{[1/N]}_{[T_1, ..., T_N]/[t_1, ..., t_N]}$ is a subset of $\mathcal{A}^{[1/N]}_{[S_1, ..., S_N]/[s_1, ..., s_N]}$

Clearly indexing hierarchy $\mathcal{A}^{[1/N]}_{[T_1, ..., T_N]/[t_1, ..., t_N]}$ inherits indexing hierarchy's $\mathcal{A}^{[1/N]}_{[S_1, ..., S_N]/[s_1, ..., s_N]}$ order.

Q.E.D
∎

**Definition 5.2:** Let $P^N = TI^{[1/N]}_{[S_1, ..., S_N]/[s_1, ..., s_N]}$ be a multi-array. Let multi-array $Q^N = TI^{[1/N]}_{[T_1, ..., T_N]/[t_1, ..., t_N]}$ be a sub-array of array $P^N$.

Let $\mathcal{A}_{P^N}$ be array $P^N$ Cartesian Extension. Let $\mathcal{P}_{P^N}$ be multi-cube's $\mathcal{A}_{P^N}$ path-set.

Let $\mathcal{P}_{Q^N/P^N} = \{ <A> = <a_1 \ldots a_N> : (a_1 \ldots a_N) \in I^{[1/N]}_{[T_1, ..., T_N]/[t_1, ..., t_N]} \}$

be a subset of path-set $\mathcal{P}_{P^N}$.

Let $\mathcal{A}_{Q^N/P^N} = \{ A \in \mathcal{A}_{P^N} : \text{there is path } <A> \in \mathcal{P}_{Q^N/P^N} \text{ such that } A \in <A> \}$ be a subset of hierarchy $\mathcal{A}_{P^N}$

We define multi-array's $Q^N$ multi-array-$P^N$-embedded Cartesian Extension, as set $\mathcal{A}_{Q^N/P^N}$ that inherits hierarchy's $\mathcal{A}_{P^N}$ order.

---

**Lemma 5.1:** Let $P^N = TI^{[1/N]}_{[S_1, ..., S_N]/[s_1, ..., s_N]}$ and $Q^N = TI^{[1/N]}_{[T_1, ..., T_N]/[t_1, ..., t_N]}$ be multi-arrays.

Let multi-array $Q^N$ be a sub-array of multi-array $P^N$.

Let $\mathcal{A}_{P^N}$ be array $P^N$ Cartesian Extension. Let $\mathcal{A}_{Q^N/P^N}$ be a multi-array's $Q^N$ multi-array $P^N$-embedded Cartesian Extension.

Then $\mathcal{A}_{Q^N/P^N} = \{ A : A = [a_1 \ldots a_M] \in \mathcal{A}_{P^N}$ and there exists node $[c_1 \ldots c_N] \in \mathcal{A}_{P^N}$ such that
  (a) $(c_1 \ldots c_N) \in I^{[1/N]}_{[T_1, ..., T_N]/[t_1, ..., t_N]}$, and
  (b) $(a_1 \ldots a_M) = (c_1 \ldots c_M)$
 $\}$

**Proof:**

Let $T_{[S_1, ..., S_N]/[s_1, ..., s_N]} : \mathcal{A}^{[1/N]}_{[S_1, ..., S_N]/[s_1, ..., s_N]} \to \mathcal{A}_{P^N}$ be an onto, one-to-one, order-

preserving map such that for $(a_1 \ldots a_M) \in \mathcal{A}^{[1/N]}_{[T_1, \ldots, T_N]/[t_1, \ldots, t_N]}$

$$T_{[S_1, \ldots, S_N]/[s_1, \ldots, s_N]}((a_1 \ldots a_M)) = [a_1 \ldots a_M] \in \mathcal{A}_{P^N}.$$

Let's assume that $A = [a_1 \ldots a_M] \in \mathcal{A}_{Q^N/P^N}$.

We have to show that node $[a_1 \ldots a_M] \in \mathcal{A}_{P^N}$ is such that there exists node $[c_1 \ldots c_N] \in \mathcal{A}_{P^N}$ such that

(a) $(c_1 \ldots c_N) \in I^{[1/N]}_{[T_1, \ldots, T_N]/[t_1, \ldots, t_N]}$, and
(b) $(a_1 \ldots a_M) = (c_1 \ldots c_M)$,

By **definition 5.2**, since $A = [a_1 \ldots a_M] \in \mathcal{A}_{Q^N/P^N}$, there exists hierarchy's $\mathcal{A}_{P^N}$ path $<A> = <c_1 \ldots c_N>$ such that $(c_1 \ldots c_N) \in I^{[1/N]}_{[T_1, \ldots, T_N]/[t_1, \ldots, t_N]}$ and $[a_1 \ldots a_M] \in <c_1 \ldots c_N>$.

Then, by **lemma 2.2**, node $[c_1 \ldots c_N] = T_{[S_1, \ldots, S_N]/[s_1, \ldots, s_N]}((c_1 \ldots c_N))$ is parsing sequence's $<A>$ terminal node.

Thus node $[c_1 \ldots c_N]$ is such that (a) $(c_1 \ldots c_N) \in I^{[1/N]}_{[T_1, \ldots, T_N]/[t_1, \ldots, t_N]}$.

By **lemma 2.2**, $[a_1 \ldots a_M] \in <A>$ means that (b) $(a_1 \ldots a_M) = (c_1 \ldots c_M)$.

Let's prove the opposite.

Let's assume that node $[a_1 \ldots a_M] \in \mathcal{A}_{P^N}$ is such that there exists node $[c_1 \ldots c_N] \in \mathcal{A}_{P^N}$ such that

(a) $(c_1 \ldots c_N) \in I^{[1/N]}_{[T_1, \ldots, T_N]/[t_1, \ldots, t_N]}$, and
(b) $(a_1 \ldots a_M) = (c_1 \ldots c_M)$,

and show that $[a_1 \ldots a_M] \in \mathcal{A}_{Q^N/P^N}$.

We have to show that there exists hierarchy's $\mathcal{A}_{P^N}$ path $<A> = <c_1 \ldots c_N>$ such that $(c_1 \ldots c_N) \in I^{[1/N]}_{[T_1, \ldots, T_N]/[t_1, \ldots, t_N]}$ and $[a_1 \ldots a_M] \in <c_1 \ldots c_N>$.

By **theorem 1.10** node $(c_1 \ldots c_N)$ is hierarchy's $\mathcal{A}^{[1/N]}_{[S_1, \ldots, S_N]/[s_1, \ldots, s_N]}$ terminal node.

By **theorem 1.13** node $T_{[S_1, ..., S_N]/[s_1, ..., s_N]}((c_1 ... c_N)) = [c_1 ... c_N]$ is hierarchy's $\mathcal{A}_{P^N}$ terminal node.

By **theorem 1.11** hierarchy $\mathcal{A}_{P^N}$ is a meta-parsing hierarchy.

By **theorems 1.2** and **1.4**, node $[c_1 ... c_N]$ uniquely defines its encompassing parsing path.

By **theorem 2.5** hierarchy $\mathcal{A}_{P^N}$ is a multi-cube.

By **lemma 2.2**, node $[c_1 ... c_N]$ uniquely defines its encompassing parsing path $<c_1 ... c_N>$.

By **definition 5.2**, path $<c_1 ... c_N> \in \mathcal{P}_{Q^N/P^N}$.

By **lemma 2.2**, since $(a_1 ... a_M) = (c_1 ... c_M)$, $[a_1 ... a_M] \in <c_1 ... c_N>$.

Thus, $[a_1 ... a_M] \in \mathcal{A}_{Q^N/P^N}$.

**Q.E.D**

■

**Lemma 5.2 :** Let $P^N = TI^{[1/N]}_{[S_1, ..., S_N]/[s_1, ..., s_N]}$ and $Q^N = TI^{[1/N]}_{[T_1, ..., T_N]/[t_1, ..., t_N]}$ be multi-arrays.

Let multi-array $Q^N$ be a sub-array of multi-array $P^N$.

Let $\mathcal{A}_{P^N}$ be array $P^N$ Cartesian Extension. Let $\mathcal{A}_{Q^N/P^N}$ be multi-array $Q^N$ multi-array $P^N$-embedded Cartesian Extension.

Then hierarchy $\mathcal{A}_{Q^N/P^N}$ is an $[T_1 ... T_N]/[t_1 ... t_N]$ indexing order hierarchy.

**Proof:**

Let $\mathcal{A}^{[1/N]}_{[S_1, ..., S_N]/[s_1, ..., s_N]}$ be an $[S_1 ... S_N]/[s_1 ... s_N]$ indexing hierarchy.
Let $\mathcal{A}^{[1/N]}_{[T_1, ..., T_N]/[t_1, ..., t_N]}$ be a $[T_1 ... T_N]/[t_1 ... t_N]$ indexing hierarchy.

Let $T_{[S_1, ..., S_N]/[s_1, ..., s_N]}$ be a map $\mathcal{A}^{[1/N]}_{[S_1, ..., S_N]/[s_1, ..., s_N]} \to \mathcal{A}_{P^N}$ such that for

$$(a_1 \ldots a_M) \in \mathcal{A}^{[1/N]}_{[S_1, \ldots, S_N]/[s_1, \ldots, s_N]}$$

$$T_{[S_1, \ldots, S_N]/[s_1, \ldots, s_N]}((a_1 \ldots a_M)) = [a_1 \ldots a_M] \equiv P^{N-M}_{[a_1, \ldots, a_M]}.$$

Let map $T_{[T_1, \ldots, T_N]/[t_1, \ldots, t_N]}$ be a map $\mathcal{A}^{[1/N]}_{[T_1, \ldots, T_N]/[t_1, \ldots, t_N]} \to \mathcal{A}_{P^N}$ such that for

$$(a_1 \ldots a_M) \in \mathcal{A}^{[1/N]}_{[T_1, \ldots, T_N]/[t_1, \ldots, t_N]}$$

$$T_{[S_1, \ldots, S_N]/[s_1, \ldots, s_N]}((a_1 \ldots a_M)) = [a_1 \ldots a_M] \equiv P^{N-M}_{[a_1, \ldots, a_M]}.$$

We have to show that map $T_{[T_1, \ldots, T_N]/[t_1, \ldots, t_N]} : \mathcal{A}^{[1/N]}_{[T_1, \ldots, T_N]/[t_1, \ldots, t_N]} \to \mathcal{A}_{Q^N/P^N}$ is an onto, one-to-one, data order preserving map.

Since multi-array $Q^N$ is a sub-array of multi-array $P^N$, multi-array $Q^N$ indexing set $I^{[1/N]}_{[T_1, \ldots, T_N]/[t_1, \ldots, t_N]}$ is a subset of multi-array $P^N$ indexing set $I^{[1/N]}_{[S_1, \ldots, S_N]/[s_1, \ldots, s_N]}$.

Therefore indexing hierarchy $\mathcal{A}^{[1/N]}_{[T_1, \ldots, T_N]/[t_1, \ldots, t_N]}$ is a subhierarchy of indexing hierarchy $\mathcal{A}^{[1/N]}_{[S_1, \ldots, S_N]/[s_1, \ldots, s_N]}$ (**lemma 5 / 1**).

Map $T_{[T_1, \ldots, T_N]/[t_1, \ldots, t_N]}$ is a subset of map $T_{[S_1, \ldots, S_N]/[s_1, \ldots, s_N]}$ and as such is a one-to-one, order-preserving map.

Thus, to prove that hierarchy is an a $[T_1 \ldots T_N]/[t_1 \ldots t_N]$ indexing order hierarchy, it is sufficient to prove that

$$T_{[T_1, \ldots, T_N]/[t_1, \ldots, t_N]}(\mathcal{A}^{[1/N]}_{[T_1, \ldots, T_N]/[t_1, \ldots, t_N]}) = \mathcal{A}_{Q^N/P^N} \text{ or,}$$

equivalently, that

$$T_{[S_1, \ldots, S_N]/[s_1, \ldots, s_N]}(\mathcal{A}^{[1/N]}_{[T_1, \ldots, T_N]/[t_1, \ldots, t_N]}) = \mathcal{A}_{Q^N/P^N}.$$

In order to prove that we will first show that $T_{[S_1, \ldots, S_N]/[s_1, \ldots, s_N]}(\mathcal{A}^{[1/N]}_{[T_1, \ldots, T_N]/[t_1, \ldots, t_N]})$ is a subset of $\mathcal{A}_{Q^N/P^N}$,

By definition, for $(a_1 \ldots a_M) \in \mathcal{A}^{[1/N]}_{[S_1, \ldots, S_N]/[s_1, \ldots, s_N]}$

$$T_{[S_1, \ldots, S_N]/[s_1, \ldots, s_N]}((a_1 \ldots a_M)) \equiv [a_1 \ldots a_M] \equiv$$

$$P^{N-M}_{[a_1, \ldots, a_M]} \equiv$$

$$\{ ((a_1, ..., a_M, a_{M+1}, ..., a_N), P^N((a_1, ..., a_M, a_{M+1}, ..., a_N))):$$
$$(a_{M+1}, ..., a_N) \in I^{[M+1/N-M]}_{[S_{M+1}, ..., S_N]/[s_{M+1}, ..., s_N]} \} \equiv$$
$$\{ ((c_1, ..., c_N), P^N((c_1, ..., c_N))):$$
$$(c_1, ..., c_N) \in \{a_1\} \times ... \times \{a_M\} \times I^{[M+1/N-M]}_{[S_{M+1}, ..., S_N]/[s_{M+1}, ..., s_N]} \}$$

Let's assume that $(a_1 ... a_M) \in \mathcal{A}^{[1/N]}_{[T_1, ..., T_N]/[t_1, ..., t_N]}$.

We will show that node $[a_1 ... a_M] \equiv P^{N-M}_{[a_1, ..., a_M]} \equiv T_{[S_1, ..., S_N]/[s_1, ..., s_N]}((a_1 ... a_M))$

is an element of $\mathcal{A}_{Q^N/P^N}$.

We first notice that since $(a_1 ... a_M) \in \mathcal{A}^{[1/N]}_{[T_1, ..., T_N]/[t_1, ..., t_N]}$
$$(a_1 ... a_M) \in I^{[1/M]}_{[T_1, ..., T_M]/[t_1, ..., t_M]}.$$

In accordance with **lemma 5 / 1**, in order to show that node $[a_1, ..., a_M] \in \mathcal{A}_{Q^N/P^N}$, we have to show that there exists node

$$[c_1 ... c_N] \equiv P^0_{[a_1, ..., a_N]} \in \mathcal{A}_{P^N} \text{ such that}$$

(a) $(c_1 ... c_N) \in I^{[1/N]}_{[T_1, ..., T_N]/[t_1, ..., t_N]}$, and
(b) $(a_1 ... a_M) = (c_1 ... c_M)$

By **Theorem 3 / 4,** set

$$\{ [a_1, ..., a_M, s_{M+1}+i] \in \mathcal{A}_{P^N} : s_{M+1}+i \in I^{M+1}_{S_{M+1}/s_{M+1}}, 1 \le i \le S_{M+1} \}$$

is set of all of node $[a_1, ..., a_M]$ children.

Since multi-array $Q^N$ is a sub-array of multi-array $P^N$, multi-array $Q^N$ indexing set

$I^{[1/N]}_{[T_1, ..., T_N]/[t_1, ..., t_N]}$ is a subset of multi-array $P^N$ indexing set $I^{[1/N]}_{[S_1, ..., S_N]/[s_1, ..., s_N]}$.

And since multi-array $Q^N$ indexing set $I^{[1/N]}_{[T_1, ..., T_N]/[t_1, ..., t_N]}$ is a subset of multi-array $P^N$

indexing set $I^{[1/N]}_{[S_1, ..., S_N]/[s_1, ..., s_N]}$, one-dimensional indexing set $I^{M+1}_{T_{M+1}/t_{M+1}}$ is a subset of one-dimensional indexing set $I^{M+1}_{S_{M+1}/s_{M+1}}$.

Therefore we can choose $1 \leq i \leq S_{M+1}$ such that $c_{M+1} = s_{M+1} + i \in I^{M+1}_{T_{M+1}/t_{M+1}}$.

Then, node $[a_1, ..., a_M, c_{M+1}]$ is node's $[a_1, ..., a_M]$ child such that

$$(a_1, ..., a_M, a_{M+1}) \in I^{[1/M+1]}_{[T_1, ..., T_{M+1}]/[r_1, ..., r_{M+1}]}$$

In this way, starting with node $[a_1, ..., a_M]$, in $N - M$ steps, we obtain node

$[a_1, ..., a_M, c_{M+1}, ..., c_N] \in \mathcal{A}_{P^N}$ such that

$$\text{node } (a_1, ..., a_M, c_{M+1}, ..., c_N) \in I^{[1/N]}_{[T_1, ..., T_N]/[t_1, ..., t_N]}.$$

Thus, by **lemma 5 / 1**, node $[a_1, ..., a_M] \equiv T_{[S_1, ..., S_N]/[s_1, ..., s_N]}((a_1 ... a_M)) \in \mathcal{A}_{Q^N/P^N}$.

We now will show the reverse, namely that if node $[a_1, ..., a_M] \in \mathcal{A}_{Q^N/P^N}$ then

$$(a_1 ... a_M) = (T_{[S_1, ..., S_N]/[s_1, ..., s_N]})^{-1}([a_1, ..., a_M]) \in \mathcal{A}^{[1/N]}_{[T_1, ..., T_N]/[t_1, ..., t_N]}.$$

By **lemma 5.1**, $[a_1, ..., a_M] \in \mathcal{A}_{Q^N/P^N}$
        **iff**

there exista node $[c_1 ... c_N] \in \mathcal{A}_{P^N}$ such that

    (a) $(c_1 ... c_N) \in I^{[1/N]}_{[T_1, ..., T_N]/[t_1, ..., t_N]}$, and
    (b) $(a_1 ... a_M) = (c_1 ... c_M)$

Since, by assumption, $[a_1, ..., a_M] \in \mathcal{A}_{Q^N/P^N}$,

this implies that $(a_1 ... a_M) \in I^{[1/M]}_{[T_1, ..., T_M]/[t_1, ..., t_M]}$.

Since, by definition, $I^{[1/M]}_{[T_1, ..., T_M]/[t_1, ..., t_M]}$ is a subset of $\mathcal{A}^{[1/N]}_{[T_1, ..., T_N]/[t_1, ..., t_N]}$,
that in turn implies that $(a_1 ... a_M) \in \mathcal{A}^{[1/N]}_{[T_1, ..., T_N]/[t_1, ..., t_N]}$.

    **Q.E.D**
■

**Lemma 5.3 :** Let $P^N = TI^{[1/N]}_{[S_1, ..., S_N]/[s_1, ..., s_N]}$ and $Q^N = TI^{[1/N]}_{[T_1, ..., T_N]/[t_1, ..., t_N]}$ be multi-arrays.

Let multi-array $Q^N$ be a sub-array of multi-array $P^N$. Let $\mathcal{A}_{Q^N/P^N}$ be multi-array $Q^N$ multi-array

$P^N$-embedded Cartesian Extension.

Then hierarchy $\mathcal{A}_{Q^N/P^N}$ is a $[T_1 \ldots T_N]/[t_1 \ldots t_N]$ multi-cube.

**Proof:**

Follows directly from **theorem 2 / 5.**

Q.E.D

∎

**Lemma 5.4 :** Let $P^N = TI^{[1/N]}_{[S_1, \ldots, S_N]/[s_1, \ldots, s_N]}$ and $Q^N = TI^{[1/N]}_{[T_1, \ldots, T_N]/[t_1, \ldots, t_N]}$ be multi-arrays.

Let multi-array $Q^N$ be a sub-array of multi-array $P^N$. Let $\mathcal{A}_{P^N}$ be multi-array $P^N$ Cartesian Extension.

Let $\mathcal{A}_{Q^N/P^N}$ be multi-array $Q^N$ multi-array $P^N$-embedded Cartesian Extension.

Then set $\{ ((a_1 \ldots a_N), P^N((a_1 \ldots a_N))) : (a_1 \ldots a_N) \in I^{[1/N]}_{[T_1, \ldots, T_N]/[t_1, \ldots, t_N]} \} \equiv$

$\{ ((a_1 \ldots a_N), Q^N((a_1 \ldots a_N))) : (a_1 \ldots a_N) \in I^{[1/N]}_{[T_1, \ldots, T_N]/[t_1, \ldots, t_N]} \}$

is mult-cube's $\mathcal{A}_{Q^N/P^N}$ data-set.

**Proof:**

Let map $T_{[T_1, \ldots, T_N]/[t_1, \ldots, t_N]}$ be a map $\mathcal{A}^{[1/N]}_{[T_1, \ldots, T_N]/[t_1, \ldots, t_N]} \to \mathcal{A}_{P^N}$ such that for

$(a_1 \ldots a_M) \in \mathcal{A}^{[1/N]}_{[T_1, \ldots, T_N]/[t_1, \ldots, t_N]}$

$T_{[S_1, \ldots, S_N]/[s_1, \ldots, s_N]}((a_1 \ldots a_M)) = [a_1 \ldots a_M] \equiv P^{N-M}_{[a_1, \ldots, a_M]}$.

We have shown that map $T_{[T_1, \ldots, T_N]/[t_1, \ldots, t_N]} : \mathcal{A}^{[1/N]}_{[T_1, \ldots, T_N]/[t_1, \ldots, t_N]} \to \mathcal{A}_{Q^N/P^N}$ is an onto, one-to-one, data order preserving map.

By **theorem 1 / 10**, indexing order hierarchy's $\mathcal{A}_{Q^N/P^N}$ data-set is

$T_{[T_1, \ldots, T_N]/[t_1, \ldots, t_N]}( I^{[1/N]}_{[T_1, \ldots, T_N]/[t_1, \ldots, t_N]} ) \equiv$

$\{ \{ ((a_1, \ldots, a_N)), P^N((a_1, \ldots, a_N))) \} : (a_1, \ldots, a_N) \in I^{[1/N]}_{[T_1, \ldots, T_N]/[t_1, \ldots, t_N]} \} \equiv$

$$\{ \ \{ \ (\ (a_1, ..., a_N)\ ), Q^N (\ (a_1, ..., a_N)\ )\ )\ \} : (a_1, ..., a_N) \in I^{[1/N]}_{[T_1, ..., T_N]/[t_1, ..., t_N]} \ \}$$

Q.E.D

∎

**Theorem 5.1 :** Let $P^N = TI^{[1/N]}_{[S_1, ..., S_N]/[s_1, ..., s_N]}$ be an $[S_1 \ ... \ S_N] / [s_1 \ ... \ s_N]$ multi-array.

Let $Q^N = TI^{[1/N]}_{[T_1, ..., T_N]/[t_1, ..., t_N]}$ be a $[T_1 \ ... \ T_N] / [t_1 \ ... \ t_N]$ sub-array of multi-array $P^N$. Let $\mathcal{A}_{Q^N}$ be array $Q^N$ Cartesian Extension. Let $\mathcal{A}_{Q^N/P^N}$ be array $Q^N$ $P^N$-embedded Cartesian Extension.

Then $\mathcal{A}_{Q^N/P^N} \approx \mathcal{A}_{Q^N}$.

**Proof:**

By **lemma 5.2**, hierarchy $\mathcal{A}_{Q^N/P^N}$ is an $[T_1 \ ... \ T_N] / [t_1 \ ... \ t_N]$ indexing order hierarchy.

By **lemma 5.3**, hierarchy $\mathcal{A}_{Q^N/P^N}$ is an $[T_1 \ ... \ T_N] / [t_1 \ ... \ t_N]$ multi-cube,

By **lemma 5.4**, ierarchy $\mathcal{A}_{Q^N/P^N}$ data-set is

$$\{ \ \{ \ (\ (a_1, ..., a_N)\ ), Q^N (\ (a_1, ..., a_N)\ )\ )\ \} : (a_1, ..., a_N) \in I^{[1/N]}_{[T_1, ..., T_N]/[t_1, ..., t_N]} \ \}$$

By **theorem 3.2**, hierarchy $\mathcal{A}_{Q^N}$ is an $[T_1 \ ... \ T_N] / [t_1 \ ... \ t_N]$ indexing order hierarchy.

By **theorem 3.3**, hierarchy $\mathcal{A}_{Q^N}$ is an $[T_1 \ ... \ T_N] / [t_1 \ ... \ t_N]$ multi-cube,

By **theorem 3.5**, hierarchy's $\mathcal{A}_{Q^N}$ data-set is

$$\{ \ \{ \ (\ (a_1, ..., a_N)\ ), Q^N (\ (a_1, ..., a_N)\ )\ )\ \} : (a_1, ..., a_N) \in I^{[1/N]}_{[T_1, ..., T_N]/[t_1, ..., t_N]} \ \}$$

Q. E. D.

∎

**Definition 5.3 :** Let $P^N = TI^{[1/N]}_{[S_1, ..., S_N]/[s_1, ..., s_N]}$ be an $[S_1 \ ... \ S_N] / [s_1 \ ... \ s_N]$ multi-array. Let $f_1$, ..., $f_N$ be type-**Q** quantizing functions of order $[T_1]/[t_1]$, ..., $[T_N]/[t_N]$ respectively. Let $_PT_Q$ be type **P** to type **Q** converter.

Let $Q^N = TI^{[1/N]}_{[T_1, ..., T_N]/[t_1, ..., t_N]}$ be an $[T_1 \ldots T_N] / [t_1 \ldots t_N]$ multi-array. Let multi-array $Q^N$ be a sub-array of an an $[S_1 \ldots S_N] / [s_1 \ldots s_N]$ multi-array $P^N$. Let $\mathcal{A}_{P^N}$ be array $P^N$ Cartesian Extension.

Let $\mathcal{A}_{Q^N / P^N}$ be multi-array $Q^N$ multi-array $P^N$-embedded Cartesian Extension.

We define quantizing multi-array $Q^N$ *locally*, in terms of type **P** quantizing functions $f_1, \ldots, f_N$ of order $[T_1]/[t_1] \ldots [T_N]/[t_N]$ resectively, and in terms of $_PT_Q$ type **P** to type **Q** converter, as quantizing multi-array's $Q^N$ multi-array $P^N$-embedded Cartesian Extension $\mathcal{A}_{P^N / Q^N}$ in terms of quantizing functions $f_1, \ldots, f_N$ of order $[T_1]/[t_1], \ldots, [T_N]/[t_N]$ resectively, and in terms of $_PT_Q$ type converter.

# VI. Quantizing Multi-Array: Computer Implementation.

## Computer Code.

The code in **Fig. 1** is a part of the working code implementation that can be viewed in its entirety at

http://www.wipo.int/patentscope/search/en/detail.jsf?docId=WO2010126783&recNum=1&tab=PCTDocuments&maxRec=1&office=&prevFilter=&sortOption=&queryString=AN%3AUS10%2F32142

**Fig. 1** meta-code – once mapped to the user-specified dimension, **N**, and then template-instantiated with the user-defined type one-dimensional interpolators $I_1 \ldots I_N$ – implements isolating recursive core within interpolation on an **N**-dimensional grid $G^N$ in terms of one-dimensional interpolators $I_1 \ldots I_N$, and a computer-implemented data-type converter, and *then* quantizing the grid's data-base as a parallel recursion.

As will be explained later in the article, isolating recursive core within interpolation on an **N**-dimensional grid $G^N$ in terms of one-dimensional interpolators $I_1 \ldots I_N$ boils down to structually uniform yet algorithm specific, *computer-implemented* mapping of one-dimensional interpolators $I_1 \ldots I_N$ to quantizing functions $I_1 \ldots I_N$ and *then* quantizing grid $G^N$ data-base in terms of quantizing functions functions $I_1 \ldots I_N$, and an appropriate type converter, as a parallel recursion.

**Fig. 2.** provides a general layout of **Fig. 1** meta-code's dimension specific template instantiation mechanism.

```
template<class X, class Y>
```

```cpp
struct rn_base_interpolator {
    .............................................
    typedef typename X Head;
    typedef typename Y Tail;
    .............................................
    template<class STRIDES>
    size_t set_strides(STRIDES &p) const {
        (p.head = tail.head.get_data_size()) *= tail.set_strides(p.tail);
        return p.head;
    }
    .............................................
    template<typename TUPLE>
    void set_argument(const TUPLE &p) const {
      head.set_argument(p.head);
      tail.set_argument(p.tail);
    }
    .............................................
    template<typename STRIDES>
    size_t get_data_offset(const STRIDES &strds) const {
        return strds.head * head.get_data_offset() +
               tail.get_data_offset(strds.tail);
    }
    .............................................
    template<class STRIDES, class ConstIterator >
    typename iterator_value<ConstIterator>::type
    interpolate(ConstIterator data, const STRIDES &strds) const {
       size_t sz = head.get_data_order(), stride = strds.head, t = 0;
       while(t < sz) {
          head.set_data(t, tail.interpolate(data, strds.tail));
          data += stride;
          ++t;
       }
       return head.interpolate();
    }
.............................................
};
template<typename U>
struct rn_base_interpolator<U, mpl::void_> {

typedef typename U Head;
typedef typename  mpl::void_ Tail;
.............................................
template<class STRIDES>
size_t set_strides(STRIDES &p) const {
  return p.head = 1;
}
.............................................
template<typename TUPLE>
void set_argument(const TUPLE &p) const {
  head.set_argument(p.head);
}
.............................................
template<typename STRIDES>
size_t get_data_offset(const STRIDES &strds) const {
      return head.get_data_offset();
}
  .............................................
```

```
template< class STRIDES, class ConstIterator >
typename iterator_value< ConstIterator >::type
interpolate(ConstIterator data, const STRIDES &strds) const {
    size_t sz = head.get_data_order(), t = 0;
    while(t < sz) {
       head.set_data(t, *data);
       ++data;
       ++t;
    }
    return head.interpolate();
}
.............................................
};
```

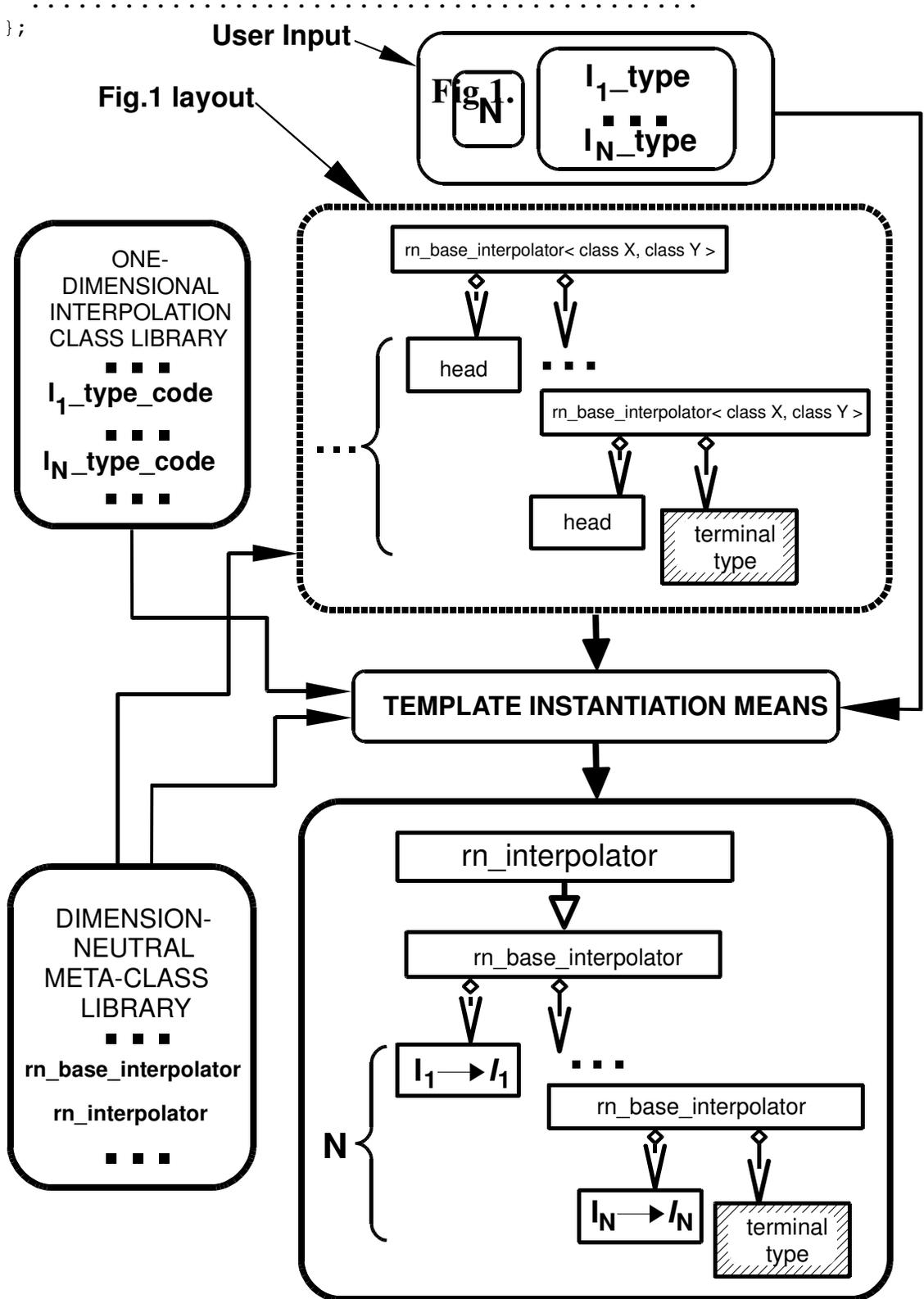

# Fig 2.

## Quantizing a Multi-Array Globally.

Let $Q^N = TI^{[1/N]}_{[S_1, ..., S_N]/[s_1, ..., s_N]}$ be a multi-array. Let $\mathcal{A}_{Q^N}$ is multi-array $Q^N$ Cartesian Extension.

Recursively defined `rn_base_interpolaton.interpolate(…)` function that performs quantizing multi-cube $Q^N$ is implemented as a parallel recursion:

A push-up part of `rn_base_interpolator.interpolate(…)` parallel recursion (see **definitions 4 / 5 and 4.6**) is implemented as a recursively embedded nested loop, thus processing multi-array $Q^N$ Cartesian Extension as an indexing order hierarchy.

A push-down, parsing, part of menber function `rn_base_interpolator.interpolate(…)` parallel recursion parses elements of hierarchy $\mathcal{A}_{Q^N}$ t contihat are contiguously stored in a computer storage device, and processess multi-array $Q^N$ Cartesian Extension elements as elements of $[CS_1 … CS_N]$-defined containment hierarchy (see theorem **theorem 3 / 4**).

Hierarchy $\mathcal{A}_{Q^N}$ strorage arrangement is implemented as followas:

First, we store multi-array $Q^N$ within a computer storage device in data-array's $Q^N$ lexicographic order :

    each of multi-array's $Q^N$ elements, $((a_1, ..., a_N), Q^N((a_1, ..., a_N)))$,

    is mapped to *storage-address / stored-value* pair

        $((A_{[a_1, ..., a_N]}, Q^N((a_1, ..., a_N))))$, wherein

    $A_{[a_1, ..., a_N]} = \&I + (a_1 - s_1) * CS_1 + ..., (a_N - s_N) * CS_N$ ,

    $\&I$ being storage-address of multi-array $Q^N$ first element.

Thus multi-array's $Q^N$ is stored as a contiguous mempry interval, $MI^N$ of size $S_1 * ... * S_N$. Once multi-array $Q^N$ has been contiguously stored within a computer storage device in multi-array's $Q^N$ lexicographic order, elements of meta-parsing hierarchy $\mathcal{A}_{Q^N}$ become *embedded* within stored multi-array $MI^N$ in a *spatial* layout that can be described a sequence of $N$ subdivision steps:

At the subdivision first step contiguous data interval $D^N$ is subdivided into $S_1$ disjoint, contiguous, $[S_1] / [s_1]$-indexed subintervals $MI^{N-1}_{[a_i]}$, $a_i \in I^1_{S_1 / s_1}$, thus each of $MI^{N-1}_{[a_i]}$ subintervals being of length $CS_1$.

Structurally, each of $MI^{N-1}_{[a_i]}$ subintervals is a computer-stored Cartesian Extension of multi-array $Q^N$ depth **1** Cartesian Projections $Q^{N-1}_{[a_i]}$, $a_i \in I^1_{[S_1]/[s_1]}$ ( **theorem 2 / 2** ).

At the subdivision second step each of contiguous subintervals $MI^{N-1}_{[a_i]}$, $a_i \in I^1_{[S_1]/[s_1]}$, of length $CS_1$ is subdivided into into $S_2$ disjoint, contiguous, $[S_2] / [s_2]$- indexed subintervals $MI^{N-2}_{[a_i, a_j]}$, $(a_i, a_j) \in I^{[1/2]}_{[S_1, S_2]/[s_1, s_2]}$, each of thus obtained subintervals being of length $CS_2$.

Structurally, each of $MI^{N-2}_{[a_i, a_j]}$ $S_1 * S_2$ subintervals is a computer-stored Cartesian Extension of multi-array $Q^N$ depth **2** Cartesian Projections $Q^{N-2}_{[a_i, a_j]}$

At the subdivision $N^{th}$ last step each of contiguous intervals subintervals $MI^1_{[a_i, ..., a_{N-1}]}$, $(a_1, ..., a_{N-1}) \in I^{[1/N-1]}_{[S_1, ..., S_{N-1}]/[s_1, ..., s_{N-1}]}$ of length $CS_{N-1}$ is subdivided into $S_N$ disjoint, $[S_N] / [s_N]$-indexed subintervals, each of thus obtained subintervals being of length $CS_N \equiv 1$.

Structurally, each of thus obtained $S_1 * ... * S_N$ subintervals is a computer-stored Cartesian Extension of multi-array $Q^N$ depth $N$ Cartesian Projections
$$Q^0_{[a_i, ..., a_N]}, (a_1, ..., a_N) \in I^{[1/N]}_{[S_1, ..., S_N]/[s_1, ..., s_N]}.$$

At this point mapping **Fig. 1** code to implementation of quantizing a multi-cube *globally,* in terms of **definitions 4.5 and 4.6,** is straightforward :

Each of depth-**i** nested `head` meta-objectsis mapped to $I_i$ quantizing function object ( **Fig. 2** ).

Each of objects' $I_i$ `.get_data_order()` calls returns $S_i$ value, thus determinig the shape of the *global* recursively embedded loop, and, within the loop, tsetting the number of arguments that quantizing function object $I_i$ takes to $S_i$, i = 1, ..., N.

In functional terms the outermost C++ call of meta-object's `rn_base_interpolator` member funcion `rn_base_interpolator.interpolate()`,

```
typename iterator_value< ConstIterator >::type
interpolate(ConstIterator data, const STRIDES &strds) const {
  size_t sz = head.get_data_order(), stride = strds.head, t = 0;
  while(t < sz) {
    head.set_data(t, tail.interpolate(data, strds.tail));
    data += stride;
    ++t;
  }
  return head.interpolate();
}
```

becomes this :

```
interpolate( &DN, [CS1 , ..., CSN] ) {
  size_t t = 0;
  while(t < S1) {
      I1.datas₁+ t = tail.interpolate( &DN − 1[s₁+ t] , [CS2 , ..., CSN] ));
      ++t;
  }
  return I1(datas₁+ 1, ..., datas₁+ S₁);
}
```

At this point we remark that the above *meta-code* snippet is a verbatim implementation of **Definition 4.4** recursion's step.

In functional terms each of the *innermost* C++ calls of meta-object's `rn_base_interpolator` member funcion `rn_base_interpolator.interpolate()`,

```
template< class STRIDES, class ConstIterator >
typename iterator_value< ConstIterator >::type
interpolate(ConstIterator data, const STRIDES &strds) const {
  size_t sz = head.get_data_order(), t = 0;
  while(t < sz) {
    head.set_data(t, *data);
    ++data;
    ++t;
  }
  return head.interpolate();
}
```

becomes :

```
interpolate(&D¹[aᵢ, ..., a_{N-1}] , [ CS_N ] ) {
  size_t t = 0;
  while(t < S_N) {

      I_N.data_{S_N + t} = * D⁰[aᵢ, ..., a_{N-1}, s_N + t];
      ++t;
  }
  return I_N(data_{S_N + 1}, ..., data_{S_N + S_N});
}
```

At this point we we remark that the above *meta-code* snippet is a verbatim implementation of **Definition 4.5** recursion's terminal step.

## Quantizing a Multi-Array Locally.

In section **V** we have defined quantizing multi-array locally ( **Definition 5.3**). In the essense, this defintion relies on mulri-array's Cartesian Extension being meta-parsing hierarchy.

We have not provided, though, *raison d'etre* for such a definition.

In fact, as we will show next, the necessity for defining and imoplementing **Definition 5. 3 arises from using local one-dimensional interpolation methods.**

Therefore, we will first describe interpolation on a multi-grid in the above described structural terms.

# VI. Interpolation on N-dimensional Grid : Definitions.

## Interpolating Function / Interpolated Function Model: a Definition

In the following sections we will take a set-theoretical view of interpolation on a grid, thus separating what is necessarily heuristic from what is not. .

As it turns out, taking such a formal view of multi-dimensional interpolation is practical to the extreme. Through approaching interpolation on a grid in a set-theoretical manner we are able to fashion *a* structural background (not necessarily the only one possible) that is suitable for discerning interpolation on a multi-grid structural bottlenecks which, without such a structural background, are elusive, difficult to put one's finger on, and – unless first identified and then eliminated – exponentially exacerbate The Curse Of Dimension.

**Definition 6.1:** We define interpolated function as a finite set-theoretical function with a numerical range.

―

**Definition 6.2:** We define interpolating function as a function that takes three variables – interpolated function domain, interpolated function argument value, and interpolated function range – and returns a numerical value.

―

**Definition 6.3:** We define one-dimensional interpolating function as an interpolation function that takes three variables – an indexed set of interpolated function known argument values, interpolated function argument value, and an indexed set of interpolated function return values at known argument values – and returns a numerical value.

―

**Definition 6.4:** Let **M** be a positive natural number. We define one-dimensional interpolating function of order **M** as an interpolation function that takes three variables – an **M-**indexed set of interpolated function known argument values, interpolated function argument value, and an **M-** indexed set of interpolated function values at ithe function's known argument values – and returns a numerical value.

―

**Definition 6.5:** Let **M** be a natural number. Let **m** be an integer number. We define one-dimensional interpolating function of order **[M] / [m]** as an interpolation function that takes three variables – an **[M] / [m]-**indexed set of interpolated function argument values, interpolated function argument value, and an **[M] / [m]-**indexed set of interpolated function known values at its known argument values – and returns a numerical value.

―

**Notation 6.1:**

▼ We will be referring to one-dimensional interpolation functions of **Definition 7 / 3** as global one-dimensional interpolation functions.

▼ We will be referring to one-dimensional interpolation functions of **Definitions 7 / 4, 7.5** and **7. 6** as local one-dimensional interpolation functions.

**Definition 6.6 :** We define interpolation as an interpolating function call.

―

# Multi-Dimensional Grid .

**Definition 6.7:** We define [$S_1$ … $S_N$] argument mesh, $M_{[S_1, …, S_N]}$, as an aggregation of **N** [$S_i$] arrays, $A_i$, such that $A_i = \{ x^i_1, …, x^i_{S_i} \}$, each forming a monotone numerical sequence, **i = 1, …, N.**

―

**Definition 6.8:** We define $[S_1 \ldots S_N] / [s_1 \ldots s_N]$ argument mesh, $M_{[S_1, \ldots, S_N]/[s_1, \ldots, s_N]}$, as an $[N]$ array of $[S_i]/[s_i]$ arrays, $A_i$, $i = 1, \ldots, N$, such that $A_i = \{ x^i_{s_i+1}, \ldots, x^i_{s_i+S_i} \}$, each forming a monotone numerical sequence.

—

Let $M_{[S_1, \ldots, S_N]}$ be an $[S_1 \ldots S_N]$ argument mesh. Let $F^N : A_1 \times \ldots \times A_N \rightarrow R^1$ be an interpolated function.

**Definition 6.9:** We define $F^N$-based grid $G^N_{[S_1, \ldots, S_N]}$ as a data set consisting of $M_{[S_1, \ldots, S_N]}$ argument mesh and

$$\mathcal{H}^N_{[S_1, \ldots, S_N]} = \{ ((i_1, \ldots, i_N), F^N(x^1_{i_1}, \ldots, x^N_{i_N})),$$

$$(i_1, \ldots, i_N) \in U^{[1/N]}_{[S_1, \ldots, S_N]}, (x^1_{i_1}, \ldots, x^N_{i_N}) \in A_1 \times \ldots \times A_N \},$$

an $[S_1 \ldots S_N]$ data-base.

—

# VII. Interpolation on a Multi-Grid – a Structural Framework.

In this section and the next we will restrict the discussion of interpolation on an $[S_1, \ldots, S_N]$ grid to the case of multi-dimensional interpolation implemented in terms of global one-dimensional interpolators.

We will consider the case of local interpolation – interpolation performed on an $[S_1, \ldots, S_N]$ grid in terms of one-dimensional interpolators $I_1, \ldots, I_N$ of order $T_1, \ldots T_N$ respectively, wherein $T_i \leq S_i$, $i = 1, \ldots, N$, in section **XI**.

Let $M_{[S_1, \ldots, S_N]}$ be a mesh.

Let $F^N = \{ ((x^1_{i_1}, \ldots, x^N_{i_N}), F^N(x^1_{i_1}, \ldots, x^N_{i_N})), (x^1_{i_1}, \ldots, x^N_{i_N}) \in A_1 \times \ldots \times A_N \}$ be an interpolated function. Let $G^N_{[S_1, \ldots, S_N]}$ be interpolated function $F^N$-based $[S_1 \ldots S_N]$ grid.

Let interpolation $I^N$ on N-dimensional grid be implemented in **N** stages, in terms of **N** one-dimensional interpolators $I_1, \ldots, I_N$ of order $[S_1], \ldots [S_N]$ respectively – each of the $I_i$ interpolators being responsible for implementing $i^{th}$ interpolation stage.

# Interpolation on a Multi-Dimensional Grid: a Standard Implementation.

**Theorem 7.2 :** An **N**-dimensional interpolation $I^N(x_1, \ldots, x_N)$ on interpolated function $F^N$, in terms of $I_1, \ldots, I_N$ one-dimensional interpolators of order $[S_1], \ldots [S_N]$ respectively, is a dimensional reduction scheme:

At the interpolation $I^N$ $1^{rst}$ interpolation stage $N - 1$ dimensional data-base

$$\mathcal{H}^{N-1}{}_{[S_1, \ldots, S_{N-1}]} = \{ \, ((i_1, \ldots, i_{N-1}), F^N(a^1{}_{i_1}, \ldots, a^{N-1}{}_{i_{N-1}}, x_N)),$$

$$(i_1, \ldots, i_{N-1}) \in U^{[1/N-1]}{}_{[S_1, \ldots, S_{N-1}]} \, \} \text{ is generated.}$$

At the interpolation $I^N$ $i^{th}$ interpolation stage $N - i$ dimensional data-base

$$\mathcal{H}^{N-i}{}_{[S_1, \ldots, S_{N-i}]} = \{ \, ((i_1, \ldots, i_{N-i}), F^N(a^1{}_{i_1}, \ldots, a^{N-i}{}_{i_{N-i}}, x_{N-i+1}, \ldots, \times x_N \,),$$

$$(i_1, \ldots, i_{N-i}) \in U^{[1/N-1]}{}_{[S_1, \ldots, S_{N-i}]} \}$$

is generated.

At the interpolation $N^{th}$ stage **0** dimensional data-base

$$\mathcal{H}^0 = \{ \, F^N(x_1, \ldots, x_N) \, \} \text{ is generated.}$$

**Proof :**.

During interpolation $I^N(x_1, \ldots, x_N)$ call:

at interpolation $I^N$ $1^{rst}$ interpolation stage:

For each of array $\mathcal{H}^N$ depth **N - 1** Cartesian projections $\mathcal{H}^{N-1}{}_{[i_1, \ldots, i_{N-1}]}$, $(i_1, \ldots, i_{N-i}) \in U^{[1/N-1]}{}_{[S_1, \ldots, S_{N-1}]}$, interpolator $I_N$ is called with

$[S_N]$ array $A_N$,

argument value $x_N$, and

[$S_N$] array $\mathcal{H}^{N-1}[i_1, ..., i_{N-1}]$ of function $F^N$ *known* values.

Thus, for each of index tuples $(i_1, ..., i_{N-1}) \in U^{[1/N-1]}[S_1, ..., S_{N-1}]$, interpolated function $F^N(\mathcal{H}^{N-1}[i_1, ..., i_{N-1}], x_N)$ value is generated.

Thus, $N-1$ dimensional data-base

$$\mathcal{H}^{N-1i}[S_1, ..., S_{N-1}] = \{((i_1, ..., i_{N-1}), F^N(\mathcal{H}^{N-1}[i_1, ..., i_{N-1}], x_N),$$

$$(i_1, ..., i_{N-i}) \in U^{[1/N-1]}[S_1, ..., S_{N-1}]\} \text{ is generated.}$$

at interpolation $I^N$ $i^{th}$ interpolation stage:

For each of array $\mathcal{H}^{N-i+1}$ depth $N-i$ Cartesian projections $\mathcal{H}^{N-i+1}[i_1, ..., i_{N-i}]$, $(i_1, ..., i_{N-i}) \in U^{[1/N-i]}[S_1, ..., S_{N-i}]$, interpolator $I_{N-i}$ is called with

[$S_{N-i+1}$] array $A_{N-i+1}$,

argument value $x_{N-i+1}$, and

[$S_{N-i+1}$] array $\mathcal{H}^{N-i+1}[i_1, ..., i_{N-1}]$ (of function $F^N$ *known* values).

Thus, for each of index-tuples $(i_1, ..., i_{N-i}) \in U^{[1/N-i]}[S_1, ..., S_{N-i}]$, interpolated function $F^N(\mathcal{H}^{N-1}[i_1, ..., i_{N-i}], x_{N-i+1}, ..., x_N)$ value is generated.

Thus, $N-i$ dimensional data-base

$$\mathcal{H}^{N-i}[S_1, ..., S_{N-i}] = \{((i_1, ..., i_{N-i}), F^N(\mathcal{H}^{N-1}[i_1, ..., i_{N-1}], x_{N-i+1}, ..., \times x_N),$$

$$(i_1, ..., i_{N-i}) \in U^{[1/N-i]}[S_1, ..., S_{N-i}]\} \text{ is generated.}$$

At interpolation $I^N$ $N^{th}$ interpolation stage:

interpolator $I_1$ is called with

[$S_1$] array $A_1$,

argument value $x_1$, and

[$S_1$] array $\mathcal{H}^1$ of function $F^N$ *known* values.

Thus, **0**-dimensional data-base $\mathcal{H}^0 = \{ F^N( x_1, ..., x_N ) \}$ is generated.

Q.E.D.

# Eliminating Redundant Data Processing in Interpolation on a Multi-Grid: the Structure and the Process.

**Theorem 7.3 :** Let $M_{[S_1, ..., S_N]}$ be a mesh.

Let $F^N = \{ ( ( x^1_{i_1}, ..., x^N_{i_N} ), F^N( x^1_{i_1}, ..., x^N_{i_N} ) ), ( x^1_{i_1}, ..., x^N_{i_N} ) \in A_1 \times ... \times A_N \}$ be an interpolated function. Let $G^N_{[S_1, ..., S_N]}$ be interpolated function $F^N$-based $[S_1 ... S_N]$ grid.

Let interpolation $I^N$ on **N**-dimensional grid be implemented in **N** stages, in terms of **N** one-dimensional interpolators $I_1, ..., I_N$ of order $[S_1], ... [S_N]$ respectively – each of the $I_i$ interpolators being responsible for implementing $i^{th}$ interpolation stage.

Within the scope of interpolation $I^N( x_1, ..., x_N )$ call $i^{th}$ interpolation stage, interpolator $I_i$ of order $[ S_i ]$ can be *redefined, in terms of its sole dependency,* as quantizing function $I_i$ of order $[ S_i ]$.

**Proof :**

Within the scope of interpolation $I^N( x_1, ..., x_N )$ call's $i^{th}$ interpolation stage:

*G*lobal interpolator $I_i$ *stage-specific* interpolation input consists of

[$S_{N-i+1}$] array $A_{N-i+1}$,

argument value $x_{N-i+1}$, and

all of array data-base $\mathcal{H}^{N-i+1}$ depth $N - i$ Cartesian projections

$\mathcal{H}^{N-i+1}{}_{[i_1, ..., i_{N-i}]}$, $(i_1, ..., i_{N-i}) \in U^{[1/N-i]}{}_{[S_1, ..., S_{N-i}]}$

That means that *within* the scope of interpolation $\mathbf{I}^N(x_1, ..., x_N)$ call's $\mathbf{i}^{th}$ interpolation stage:

(a) Interpolator's $\mathbf{I}_i$ first-and-second argument values – array $\mathbf{A}_{N-i+1}$ and interpolated function argument value $x_i$ – remain constant.

(b) Interpolator's $\mathbf{I}_i$ third argument values – data-base $\mathcal{H}^{N-i+1}$ depth $N - i$ Cartesian Projections – vary.

Thus, *within* the scope of interpolation $\mathbf{I}^N(x_1, ..., x_N)$ call's $\mathbf{i}^{th}$ interpolation stage, interpolator's $\mathbf{I}_i$ first-and-second argument values constancy provides a formal ground for *redefining of* interpolator $\mathbf{I}_i$ of order $[\mathbf{S}_i]$ in terms of its sole dependency on its third argument value – as a quantizing function $I_i$ of order $[\mathbf{S}_i]$ as follows:

for $(i_1, ..., i_{N-i}) \in U^{[1/N-1]}{}_{[S_1, ..., S_{N-1}]}$

$$I_i(\mathcal{H}^{N-i+1}{}_{[i_1, ..., i_{N-i}]}) = \mathbf{I}_i(A_i, x_i, \mathcal{H}^{N-i+1}{}_{[i_1, ..., i_{N-i}]})$$

**Q.E.D.**

In practical terms, though, within the scope of interpolation $\mathbf{I}^N(x_1, ..., x_N)$ call $\mathbf{i}^{th}$ interpolation stage, a formal redefinition of interpolator $\mathbf{I}_i$ as quantizing function $I_i$, by dint of being formal, does not affect the way interpolator $\mathbf{I}_i$ *is implemented* : quantizing function $I_i$ and interpolator $\mathbf{I}_i$ still share the same set of instructions.

A mere possibility of such redefinition does not provide an impetus for as much as writing it down.

It is *implementing* quantizing function $I_i$ as a function, though, that does provide a powerful reason for the redefinition.

Within the scope of interpolation $I^N(x_1, ..., x_N)$ call $i^{th}$ interpolation stage, interpolator $I_i$ first-and-second argument values have to be processed. Therefore, in order to *implement* quantizing function $I_i$ as a function that that processes interpolator $I_i$ third argument values only, interpolator $I_i$ first-and-second argument values must be pre-processed by *suitably modified* interpolator $I_i$ instructions set before any of quantizing function $I_i(\mathcal{H}^{N-i+1}[i_1, ..., i_{N-i}])$ calls are made.

Once it's done

**Theorem 7.4 :** Within the scope of each of interpolation $I^N(x_1, ..., x_N)$ call $i^{th}$ stages, the number of instances of interpolator $I_i$ first-and-second argument values being processed is reduced from $S_1 \times ... \times S_{N-i}$ to $1$.

**Proof :** Obvious.

**Theorem 7.5 :** Within the scope of interpolation $I^N(x_1, ..., x_N)$ call, by implementing each of interpolators $I_i$ ($i = 1, ..., N$) as quantizing function $I_i$, within interpolation $I^N(x_1, ..., x_N)$ call, redundant data processing is eliminated.

**Proof :**

Within the scope of interpolation $I^N(x_1, ..., x_N)$ call, all data processing is done locally. Q.E.D.

**Programming notice 7.1 :** Implementing each of interpolator $I_i$ ($i = 1, ..., N$) as quantizing function $I_i$ can be achieved, for example, by implementing each of interpolators $I_i$ as a code-partitioned instruction set.

# VIII. Reducing Interpolation on Multi-Grid to Quantizing Grid Data-Base as a Recursion.

Let $M_{[S_1, ..., S_N]}$ be a mesh.

Let $F^N = \{ ((x^1_{i_1}, ..., x^N_{i_N}), F^N(x^1_{i_1}, ..., x^N_{i_N})), (x^1_{i_1}, ..., x^N_{i_N}) \in A_1 \times ... \times A_N \}$ be an interpolated function. Let $G^N_{[S_1, ..., S_N]}$ be interpolated function $F^N$-based $[S_1 ... S_N]$ grid.

Let interpolation $I^N$ on N-dimensional grid be implemented in **N** stages, in terms of **N** one-dimensional interpolators $I_1, ..., I_N$ of order $[S_1], ... [S_N]$ respectively – each of the $I_i$ interpolators being responsible for implementing $i^{th}$ interpolation stage.

The previous section's *interpolation-stage-by-interpolation-stage* approach to o implementing interpolators $I_i$ as quantizing functions $I_N$ provides basis for eliminating redundant data processing during interpolation function $I^N(x_1, ..., x_N)$ single call.

In this section we modify the previous section's *interpolation-stage-by-interpolation-stage* approach to implementing interpolators $I_i$ as quantizing functions $I_N$ globally.

**Corollary 8 / 1:** Within the scope of interpolation $I^N(x_1, ..., x_N)$ *single call* – once we pre-process ***all*** of interpolators $I_i$ respective first and second argument values before ***any*** of quantizing functions $I_j$ calls are made – the remaining part of *iteratively* processing interpolation call $I^N(x_1, ..., x_N)$ consists of the following steps**:**

During interpolation $I^N(x_1, ..., x_N)$ call:

    At interpolation $I^N$ $1^{rst}$ interpolation stage:

    For each of array $\mathcal{H}^N$ depth **N - 1** Cartesian projections $\mathcal{H}^{N-1}_{[i_1, ..., i_{N-1}]}$, $(i_1, ..., i_{N-i}) \in U^{[1/N-1]}_{[S_1, ..., S_{N-1}]}$, quantizing function $I_N$ is called with

        array $\mathcal{H}^{N-1}_{[i_1, ..., i_{N-1}]}$ array of function $F^N$ *known* values.

Thus, for each of index tuples $(i_1, \ldots, i_{N-1}) \in U^{[1/N-1]}[S_1, \ldots, S_{N-1}]$, interpolated function $F^N(\mathcal{H}^{N-1}[i_1, \ldots, i_{N-1}], x_N)$ value is generated.

Thus, $N-1$ dimensional data-base

$$\mathcal{H}^{N-1} = \{((i_1, \ldots, i_{N-1}), F^N(\mathcal{H}^{N-1}[i_1, \ldots, i_{N-1}], x_{N-i+1}, \ldots, \times x_N),$$

$$(i_1, \ldots, i_{N-i}) \in U^{[1/N-1]}[S_1, \ldots, S_{N-1}]\} \text{ is generated.}$$

At interpolation $I^N$ $i^{th}$ interpolation stage:

For each of array $\mathcal{H}^{N-i+1}$ depth $N-i$ Cartesian projections $\mathcal{H}^{N-i+1}[i_1, \ldots, i_{N-i}]$, $(i_1, \ldots, i_{N-i}) \in U^{[1/N-i]}[S_1, \ldots, S_{N-i}]$, quantizing function $I_{N-i}$ is called with

$$\mathcal{H}^{N-i+1}[i_1, \ldots, i_{N-1}] \text{ array (of function } F^N \text{ known values).}$$

Thus, for each of index-tuples $(i_1, \ldots, i_{N-i}) \in U^{[1/N-i]}[S_1, \ldots, S_{N-i}]$, interpolated function $F^N(\mathcal{H}^{N-1}[i_1, \ldots, i_{N-i}], x_{N-i+1}, \ldots, x_N)$ value is generated.

Thus, $N-i$ dimensional data-base

$$\mathcal{H}^{N-i} = \{((i_1, \ldots, i_{N-i}), F^N(\mathcal{H}^{N-1}[i_1, \ldots, i_{N-1}], x_{N-i+1}, \ldots, \times x_N),$$

$$(i_1, \ldots, i_{N-i}) \in U^{[1/N-i]}[S_1, \ldots, S_{N-i}]\} \text{ is generated.}$$

At interpolation $I^N$ $N^{th}$ interpolation stage:

interpolator $I_1$ is called with array $\mathcal{H}^1$ array of function $F^N$ *known* values.

Thus, value $F^N(x_0, \ldots, x_N)$ is generated.

We now are ready to prove that

**Theorem 8.1**: Within the scope of interpolation $I^N(x_1, \ldots, x_N)$ call – once we pre-process *all* of interpolators $I_i$ ($i = 1, \ldots, N$) respective first and second argument values before *any* of quantizing functions $I_j$ (j = 1, ..., N) calls are made – processing grid $G^N$ [$S_1$ ... $S_N$] data-base, either *iteratively*, as it has been described in **corollary** 8 / 1, or by quantizing grid $G^N$ [$S_1$ ... $S_N$] data-base, in term of quantizing functions $I_1$ ... $I_N$ of order [ $S_1$ ] ... [ $S_N$ ] respectively, as a recursion – will generate identical output.

**Proof :**

Proof is by induction.

In case of interpolation in the dimension one interpolation on one-dimensional [$S_1$] grid $G^1$, either iteratively or as a recursion, is a one-step process accomplished by calling quantizing function $I_1$ of [$S_1$] order with [$S_1$] data-base $\mathcal{H}^1$ of function $F^N$ *known* values. In both cases, an identical value $I_1(\mathcal{H}^1)$ is generated.

We now assume that processing an N – 1 dimensional grid's data-base, either *iteratively*, as it has been described in **corollary 10 / 1,** or by quantizing grid $G^N$ data-base *as a recursion* – in terms of a shared set of quantizing functions -- generates the same output value.

To prove the theorem for the dimension N we now break grid $G^N$ into its $S_1$ sub-grids $G^N_i$ by reducing grid $G^N$ mesh and breaking grid $G^N$ data-base $\mathcal{H}^N$ into $S_1$ of its depth 1 Cartesian Projections $\mathcal{H}^{N-1}_{[i]}$, , $i \in I^1_{[S_1]}$ :

$$\mathcal{H}^{N-1}_{[i]} = \{ ((i, i_2, ..., i_N), F^N(x^1_i, x^2_{i_2}, ..., x^N_{i_N})),$$

$$(i_2, ..., i_N) \in U^{[2/N-1]}_{[S_2, ..., S_N]}, (x^1_{i_2}, ..., x^N_{i_N}) \in A_2 \times ... \times A_N \}$$

$$= \{ ((i_2, ..., i_N), F^{N-1}_i(x^2_{i_2}, ..., x^N_{i_N})),$$

$$(i_2, ..., i_N) \in U^{[2/N-1]}_{[S_2, ..., S_N]}, (x^1_{i_2}, ..., x^N_{i_N}) \in A_2 \times ... \times A_N \},$$

$$F^{N-1}_i(x^2_{i_2}, ..., x^N_{i_N}) = F^N(x^1_i, x^2_{i_2}, ..., x^N_{i_N}).$$

By the inductive assumption interpolating on [ $S_2$ , ..., $S_N$ ] data-bases $\mathcal{H}^{N-1}_{[i]}$ , either iteratively or as a recursion, in terms of a shared set of quantizing functions, will generate the same [$S_1$] array **F** of type-P values.

In case of iterative interpolation, the results of interpolating on $\mathcal{H}^{N-1}_{[s_1+i]}$ data-base in term of quantizing functions $I_2$ ... $I_N$ of order [ $S_2$ ] ... [ $S_N$ ] respectively, will be an **[$S_1$] array F** of interpolated function $F^{N-1}_i(x_2, \ldots, x_N)$ values. $= F^N(x^1_i, x_2, \ldots, x_N)$ values respectively $(i = 1, \ldots, S_1)$.

By definition, $F^{N-1}_i(x_2, \ldots, x_N) = F^N(x^1_i, x_2, \ldots, x_N)$

Thus, the last step of interpolation $I^N(x_1, \ldots, x_N)$ call on grid $G^N$ – either iteratively or as a recursion – is performed by quantizing function $I_1(Q)$ call.

**Q.E.D.**

**Programming notice 8.1 :** Within the scope of interpolation $I^N(x_1, \ldots, x_N)$ single call – pre-processing *all* of one-dimensional interpolators' $I_i$ constant first and second argument values $(i = 1, \ldots, N)$ before *any* of interpolators $I_j$ third argument values are processed $(j = 1, \ldots, N)$ can be accomplished, as it is illustrated by the accompanying code below, by implementing interpolation $I^N$ in terms of one-dimensional interpolators $I_1, \ldots, I_N$ as an object-within-an-objects scheme.

The working code implementation od the above arrangement can be viewed in its entirety at

http://www.wipo.int/patentscope/search/en/detail.jsf?docId=WO2010126783&recNum=1&tab=PCTDocuments&maxRec=1&office=&prevFilter=&sortOption=&queryString=AN%3AUS10%2F32142

# IX. Quantizing a Multi-Array Locally: Performance Benefits.

**I.** Through the author-extended C++ template-instantiation mechanism (not shown here), The Code can be instantiated to interpolate in any number of dimensions.

**II.** Through **t**he author-extended C++ template-instantiation mechanism , The Code can be adopted to interpolate in terms of any combination of local and global one-dimensional interpolation algorithms.

**III.** The Code implements redundant input processing elimination scheme.

**IV. (a)** The Code implements redundant data-parsing elimination scheme.
  **(b)** The Code implements redundant overhead elimination scheme.

**V.** The Code redundant stack grows elimination scheme.

# XI. Sample Test Data.

Below is sample test data we have obtained by running software that implements the above described arrangements.

# TEST METHOD:

To test an interpolation method against a benchmark function, the benchmark function's values are used to form a data grid.

At an argument node

    (a) The benchmark function is called,

    (b) The interpolation method being tested is performed on thus

    created grid, and

    (c) The outputs (a) and (b) are compared.

# HARDWARE:

# An HP laptop:

    Two AMD Phenom II N620 Dual-Core Processors

    4GB of memory

    500GB 7200RPM hard drive

# SOFTWARE:

    64-bit Windows 7

    MS Visual C++ 2008 Express Edition

## INTERPOLATION IN THE DIMENTION 6

## R6 Benchmark Function :

```
{
return log(sqrt(h0 * sqrt(log(h1)) * h8) + h7 * h9 - exp(sin(h2) *
sin(3 * h3)) + sqrt(log(h3 * h4) * sqrt(h5)) + h6 * sinh(h7 + 12));
}
```

## RATIONAL-POLYNOMIAL INTERPOLATION

### At 4 nearest points in each dimension

Interpolation Speed: 1.5 sec

| Grid Spacing | Interpolation Precision |
|---|---|
| 0.025 | .000001 |
| 0.25 | .000001 |
| 0.5 | .000001 |
| 1.0 | .00001 |
| 1.5 | .00001 |
| 2.0 | .0001 |

### At 5 nearest points in each dimension

Interpolation Speed : 15 sec

| Grid Spacing | Interpolation Precision |
|---|---|
| 0.025 | .000001 |
| 0.25 | .000001 |
| 0.5 | .000001 |
| 1.0 | .000001 |
| 1.5 | .000001 |
| 2.0 | .000001 |
| 2.5 | .00001 |
| 3.0 | .00001 |
| 4.0 | .00001 |
| 5.0 | .0001 |

### POLYNOMIAL INTERPOLATION

### At 4 nearest points in each dimension

Interpolation Speed: 1.5 sec

| Grid Spacing | Interpolation Precision |
|---|---|
| .025 | .000001 |
| .25 | .000001 |
| .5 | .000001 |
| 1.0 | .00001 |
| 1.5 | .0001 |
| 2.0 | .0001 |

## At 5 nearest points in each dimension

Interpolation Speed: 15 sec

| Grid Spacing | Interpolation Precision |
|---|---|
| 0.025 | .0000000000001 |
| 0.25 | .0000000001 |
| 0.5 | .000000001 |
| 1.0 | .0000001 |
| 1.5 | .000001 |
| 2.0 | .000001 |
| 2.5 | .00001 |
| 3.0 | .00001 |
| 4.0 | .00001 |
| 5.0 | .0001 |

# Appendix: Notation. Definitions.

## 0. Functions.

We will be using the term function as a reference to Lobachevsky's set-theoretical function.

## I. Indexing Sets

**Notation I.1:** Let $N$ be a positive natural number. Let $S_1 \ldots S_N$ be $N$ positive natural numbers. Let $s_1 \ldots s_N$ be $N$ integer numbers.

Below, we use * as a place holder.

▼ We denote set $\{ 1, 2, \ldots, S_i \}$ as $*^i_{S_i}$ indexing sets.

▼ We denote sets $\{ 1+s_i, 2+s_i, \ldots, S_i + s_i \}$ as $*^i_{S_i / s_i}$ indexing set.

▼ We denote Cartesian product $*^1_{S_1} \times \ldots \times *^N_{S_N}$ as $*^{[1/N]}_{[S_1, \ldots, S_N]}$ indexing set.

▼ We denote Cartesian product $*^1_{S_1 / s_1} \times \ldots \times *^N_{S_N / s_N}$ as $*^{[1/N]}_{[S_1, \ldots, S_N] / [s_1, \ldots, s_N]}$ indexing set.

▼ We denote Cartesian sub-product $*^K_{S_K} \times \ldots \times *^L_{S_L}$ of Cartesian Product $*^1_{S_1} \times \ldots \times *^N_{S_N}$ as $*^{[K/L-K]}_{[S_K, \ldots, S_L]/[s_L, \ldots, s_M]}$.

▼ We denote Cartesian sub-product $*^K_{S_K} \times \ldots \times *^L_{S_L}$ of Cartesian Product $*^1_{S_1} \times \ldots \times *^N_{S_N}$ as $*^{[K/L-K]}_{[S_K, \ldots, S_L]}$.

▼ We denote Cartesian sub-product $*^K_{S_K/s_K} \times \ldots \times *^L_{S_L/s_L}$ of Cartesian Product $*^1_{S_1/s_1} \times \ldots \times *^N_{S_N/s_N}$ as $*^{[K/L-K]}_{[S_K, \ldots, S_L]/[s_L, \ldots, s_M]}$.

For example:

$I^{[1/N]}_{[S_1, \ldots, S_N]}$ would be an $[S_1 \ldots S_N]$-shaped indexing set.

$U^{[1/N]}_{[S_1, \ldots, S_N]/[s_1, \ldots, s_N]}$ would be an $[s_1 \ldots s_N]$-shifted $[S_1 \ldots S_N]$-shaped indexing set.

$T^{[4/5]}_{[S_4, \ldots, S_8]/[s_4, \ldots, s_8]}$ would be a Cartesian sub-product $T^4_{S_4/s_4} \times \ldots \times T^8_{S_8/s_8}$

of a Cartesian sub-product $T^{[1/N]}_{[S_1, \ldots, S_N]/[s_1, \ldots, s_N]}$

**Notation I.2:** Let $M < N$ be a positive natural numbers. Let $S_1 \ldots S_N$ be positive natural numbers. Let $s_1 \ldots s_N$ be integer numbers. Let $U^{[1/N]}_{[S_1, \ldots, S_N]/[s_1, \ldots, s_N]}$, $U^{[1/M]}_{[S_1, \ldots, S_M]/[s_1, \ldots, s_M]}$, and $U^{[M+1/N-M]}_{[S_{M+1}, \ldots, S_N]/[s_{M+1}, \ldots, s_N]}$ be indexing sets.

▼ Within the context of $U^{[1/N]}_{[S_1, \ldots, S_N]/[s_1, \ldots, s_N]}$ indexing set will be referring to $U^{[1/M]}_{[S_1, \ldots, S_M]/[s_1, \ldots, s_M]}$ and $U^{[M+1/N-M]}_{[S_{M+1}, \ldots, S_N]/[s_{M+1}, \ldots, s_N]}$ indexing sets as orthogonal indexing sets.

## II. Indexed sets.

**Definition II.1:** Let **Q** be a set.

▼ We define set **Q** paired with $\{ Q \times Q \} \setminus \{ (a, a) : a \in Q \}$ relationship as indexable set.

▼ We denote indexing set $*^{[1/N]}_{[S_1, \ldots, S_N]/[s_1, \ldots, s_N]}$ lexicographic-order as

$<_{[S_1, \ldots, S_N]/[s_1, \ldots, s_N]}$.

**Notation II.2 :** Let **Q** be a meta-indexable set. Let **T:** $\mathbf{I}^{[1/N]}_{[S_1, ..., S_N]/[s_1, ..., s_N]} \rightarrow \mathbf{Q}$ be a map such that

(a) **T** is a one-to-one map, and

(b) $\mathbf{T}(\mathbf{I}^{[1/N]}_{[S_1, ..., S_N]/[s_1, ..., s_N]}) = \mathbf{Q}$.

▼ We will be referring to set **Q** paired with map **T** embedded order as
$[S_1, ..., S_N]$ i / $[s_1, ..., s_N]$ indexed set.

▼ We will be referring to map **T** as set $\mathbf{Q}\ [S_1, ..., S_N] / [s_1, ..., s_N]$ indexing map.

Let set **Q** be an $[S_1, ..., S_N]$ indexable set.
Let map map **T:** $\mathbf{I}^{[1/N]}_{[S_1, ..., S_N]/[s_1, ..., s_N]} \rightarrow \mathbf{Q}$ be set **Q** indexing map.

Let set **Q** order, $<^T_{[S_1, ..., S_N]/[s_1, ..., s_N]}$, be defined as follows.

For **a, b** ∈ **Q** $\mathbf{a} <^T_{[S_1, ..., S_N]/[s_1, ..., s_N]} \mathbf{b}$ iff $\mathbf{T}^{-1}(\mathbf{a}) <_{[S_1, ..., S_N]/[s_1, ..., s_N]} \mathbf{T}^{-1}(\mathbf{b})$

**Notation II.3 :**

▼ We will be referring to set $\mathbf{Q} <^T_{[S_1, ..., S_N]/[s_1, ..., s_N]}$ order as map **T**-indexing order.

▼ Where no ambiguity arises will be referring to set **Q** paired with $<^T_{[S_1, ..., S_N]/[s_1, ..., s_N]}$ order as $[S_1, ..., S_N] / [s_1, ..., s_N]$ -indexed set.

# III. Arrays.

**Definition III.1 :** We define multi-array as a function whose domain is an indexing set.

**Definition III.2:** Let **A** and **B** be arrays. We define array **A** as a subarray of array **B** if array **A** is a subset of array **B**..

**Notation III / 1 :** ▼ We will reserve notation $*_*^{[1/N]}{}_{[S_1, ..., S_N]/[s_1, ..., s_N]}$
for arrays that have indexing sets $*^{[1/N]}_{[S_1, ..., S_N]/[s_1, ..., s_N]}$ as their domain.

For example :

declaring $AU^{[1/N]}_{[S_1, ..., S_N] / [s_1, ..., s_N]}$ to be an array would mean that array $AU^{[1/N]}_{[S_1, ..., S_N] / [s_1, ..., s_N]}$ has indexing set $U^{[1/N]}_{[S_1, ..., S_N] / [s_1, ..., s_N]}$ as its domain.

Let $AU^{[1/N]}_{[S_1, ..., S_N] / [s_1, ..., s_N]}$ and $AU^{[1/N]}_{[T_1, ..., T_N] / [t_1, ..., t_N]}$ be arrays.
Let indexing set $U^{[1/N]}_{[S_1, ..., S_N] / [s_1, ..., s_N]}$ be a subset of indexing set $U^{[1/N]}_{[T_1, ..., T_N] / [t_1, ..., t_N]}$.

**Notation III.2 :**

▼ We use an <u>overlapping</u> 'AU' notation to indicate that array $AU^{[1/N]}_{[S_1, ..., S_N] / [s_1, ..., s_N]}$ is a sub-array of array $AU^{[1/N]}_{[T_1, ..., T_N] / [t_1, ..., t_N]}$.

▼ We will be referring to $*_*^{[1/N]}{}_{[S_1, ..., S_N] / [s_1, ..., s_N]}$ arrays as $[S_1, ..., S_N] / [s_1, ..., s_N]$ arrays.

▼ We will be referring to $*_*^{[1/N]}{}_{[S_1, ..., S_N] / [s_1, ..., s_N]}$ arrays as **N-dimensional** arrays.

▼

A shorthand: an expression like like

'Let $AU^{[1/N]}_{[S_1, ..., S_N] / [s_1, ..., s_N]}$ be an array'

should be read as a shorthand for

'Let **N** be a positive natural number. Let $S_1 ... S_N$ be **N** positive natural numbers.
Let $s_1 ... s_N$ be **N** integer numbers. Let $U^{[1/N]}_{[S_1, ..., S_N] / [s_1, ..., s_N]}$ be an indexing set.
Let $AU^{[1/N]}_{[S_1, ..., S_N] / [s_1, ..., s_N]}$ be an array with indexing set $U^{[1/N]}_{[S_1, ..., S_N] / [s_1, ..., s_N]}$) as its domain.'

There will be other shorthand-ed statements analogous to the above that, we hope, will be readily recognizable and easily parsed.

# IV. Cartesian Strides.

Let $[*_1, ..., *_N]$ be a set of positive natural numerals.

▼ We denote $[*_1, ..., *_N]$-derived set of Cartesian Strides as $[C^*_1, ..., C^*_N]$, where

$$C^*_N \equiv 1,$$
$$...$$
$$C^*_i \equiv *_{i+1} * C^*_{i+1} = *_{i+1} * *_{i+2} * ... * *_N,$$
$$...$$

$$C^*_1 \equiv {}^*_2 * C^*_2 = {}^*_2 * {}^*_3 * \ldots * {}^*_N.$$

## V. Types.

**Definition V.1:** Let **Q** be a set. We define type **Q** as set **Q** paired with $\{ \mathbf{Q} \times \mathbf{Q} \} \setminus \{ (s,s) : s \in \mathbf{Q}\}$ relationship.

▬

**Notation V/1:** Let **P** be a type.

▼ Elements of set **P** will be referred to as elements of type **P.**

▼

**Lemma V.1 :** A subset of a type is a type**.**

**Proof:**

Let **Q**, **R** be types. Let set **Q** be a subset of set **R**.

Set $\{ \mathbf{Q} \times \mathbf{Q} \} \setminus \{ (s,s) : s \in \mathbf{Q}\}$ is a subset of set $\{ \mathbf{R} \times \mathbf{R} \} \setminus \{ (s,s) : s \in \mathbf{R}\}$.

Since set **Q** is paired with $\{ \mathbf{R} \times \mathbf{R} \} \setminus \{ (s,s) : s \in \mathbf{R}\}$ relationship, set **Q** is paired with $\{ \mathbf{Q} \times \mathbf{Q} \} \setminus \{ (s,s) : s \in \mathbf{Q}\}$ relationship as well.

**Q.E.D.**

■

**Definition V.2** Let **P**, **Q** be types. We define map $_\mathbf{P}\mathbf{T}_\mathbf{Q} : \mathbf{P} \to \mathbf{Q}$ as type **P** to type **Q** converter if $_\mathbf{P}\mathbf{T}_\mathbf{Q}$ is a one-to-one map.

▬

**Corollary :** The above definitions of type and type converter are minimalistic enough to be non-heuristic, yet powerful enough to guarantee, for example, that there is no valid converter that maps real numbers to integers.

**Notation V.2:** : Let **P, Q** be types. Let type **P** be a subset of type **Q.**

▼ Elements of set **P** may be referred to as elements of type **Q.**

▼

## VI. Hierarchies.

**Definition VI.1:** Let **Q** be a set. Let binary order **>** be a transitive, anti-symmetric subset of set

$Q \times Q \setminus \{ (x, x) : x \in Q \}$.

We define strictly-ordered hierarchy $Q$ as set $Q$ paired with $>$ binary order.

---

**Lemma VI.1 :** Let $Q$ be $>$-strictly-ordered hierarchy. Let $A, B \in Q$. Let $A > B$. Then $A \neq B$.

**Proof:**

    By definition, set $Q$ inequality relationship, $\neq, \equiv Q \times Q \setminus \{ (x, x) : x \in Q \}$.

    By definition, $>$–binary relationship is a subset of set $Q \times Q \setminus \{ (x, x) : x \in Q \}$.

**Q. E. D**.

■